\documentstyle[12pt,psfig]{article}
\makeatletter
\def\@cite#1#2{{[{#1}]\if@tempswa\typeout
{IJCGA warning: optional citation argument
ignored: `#2'} \fi}}
\newcount\@tempcntc
\def\@citex[#1]#2{\if@filesw\immediate\write\@auxout{\string\citation{#2}}\fi
 \@tempcnta\z@\@tempcntb\m@ne\def\@citea{}\@cite{\@for\@citeb:=#2\do
   {\@ifundefined
    {b@\@citeb}{\@citeo\@tempcntb\m@ne\@citea\def\@citea{,}{\bf ?}\@warning
     {Citation `\@citeb' on page \thepage \space undefined}}%
    {\setbox\z@\hbox{\global\@tempcntc0\csname b@\@citeb\endcsname\relax}%
    \ifnum\@tempcntc=\z@ \@citeo\@tempcntb\m@ne
   \@citea\def\@citea{,}\hbox{\csname b@\@citeb\endcsname}%
     \else
      \advance\@tempcntb\@ne
      \ifnum\@tempcntb=\@tempcntc
      \else\advance\@tempcntb\m@ne\@citeo
      \@tempcnta\@tempcntc\@tempcntb\@tempcntc\fi\fi}}\@citeo}{#1}}
\def\@citeo{\ifnum\@tempcnta>\@tempcntb\else\@citea\def\@citea{,}%
  \ifnum\@tempcnta=\@tempcntb\the\@tempcnta\else
   {\advance\@tempcnta\@ne\ifnum\@tempcnta=\@tempcntb \else
\def\@citea{--}\fi
   \advance\@tempcnta\m@ne\the\@tempcnta\@citea\the\@tempcntb}\fi\fi}
\makeatother

\def\boxit#1{\leavevmode\thinspace\hbox{\vrule\vtop{\vbox{\hrule%
        \vskip3pt\kern1pt\hbox{\vphantom{\bf/}\thinspace\thinspace%
        {\bf#1}\thinspace\thinspace}}\kern1pt\vskip3pt\hrule}\vrule}%
        \thinspace}
\def\Boxit#1{\noindent\vbox{\hrule\hbox{\vrule\kern3pt\vbox{
        \advance\hsize-7pt\vskip-\parskip\kern3pt\bf#1
        \hbox{\vrule height0pt depth\dp\strutbox width0pt}
        \kern3pt}\kern3pt\vrule}\hrule}}

\newcommand{\Hh}{\lower1.2ex\hbox{$\stackrel{\textstyle
H}{\footnotesize\sim}$}}
\newcommand{\Hho}{\lower1.2ex\hbox{$\stackrel{\textstyle
H_1}{\footnotesize\sim}$}}
\newcommand{\Hhw}{\lower1.2ex\hbox{$\stackrel{\textstyle
H_2}{\footnotesize\sim}$}}
\newcommand{\h}{\lower1.2ex\hbox{$\stackrel{\textstyle
h}{\footnotesize\sim}$}}

\newcommand{\gsim}{\lower.7ex\hbox{$\;\stackrel{\textstyle>}{\sim}\;$}}
\newcommand{\lsim}{\lower.7ex\hbox{$\;\stackrel{\textstyle<}{\sim}\;$}}
\newcommand{\be}{\begin{equation}}
\newcommand{\ee}{\end{equation}}
\newcommand{\bea}{\begin{eqnarray}}
\newcommand{\eea}{\end{eqnarray}}
\newcommand{\SUSY}{\makebox[1.25cm][l]{$\line(4,1){35}$\hspace{-1.15cm}{SUSY}}}

\def\simlt{\stackrel{<}{{}_\sim}}

\def\baselinestretch{1}
\begin{document}
%%%%%%%%%%%%%%%%%%%%%%%%%%% subequations.sty %%%%%%%%%%%%%%%%%%%%%%%%
\catcode`@=11
\newtoks\@stequation
\def\subequations{\refstepcounter{equation}%
\edef\@savedequation{\the\c@equation}%
  \@stequation=\expandafter{\theequation}%   %only want \theequation
  \edef\@savedtheequation{\the\@stequation}% % expanded once
  \edef\oldtheequation{\theequation}%
  \setcounter{equation}{0}%
  \def\theequation{\oldtheequation\alph{equation}}}
\def\endsubequations{\setcounter{equation}{\@savedequation}%
  \@stequation=\expandafter{\@savedtheequation}%
  \edef\theequation{\the\@stequation}\global\@ignoretrue

\noindent}
\catcode`@=12
%%%%%%%%%%%%%%%%%%%%%%%%%%%%%%%%%%%%%%%%%%%%%%%%%%%%%%%%%%%%%%%%%%%%%
\begin{titlepage}

\title{{\bf Low-scale supersymmetry breaking:
effective description,
electroweak breaking and phenomenology}}
\vskip3in
\author{  {\bf A. Brignole$^{1}$\footnote{\baselineskip=16pt E-mail 
address: {\tt
andrea.brignole@pd.infn.it}}},
{\bf J.A. Casas$^{2,3}$\footnote{\baselineskip=16pt E-mail address: {\tt
alberto@makoki.iem.csic.es}}},
{\bf J.R. Espinosa$^{3,4}$\footnote{\baselineskip=16pt E-mail address: {\tt
espinosa@makoki.iem.csic.es}}} and 
{\bf I. Navarro$^{2,5}$\footnote{\baselineskip=16pt E-mail address: {\tt
ignacio.navarro@durham.ac.uk}}}
\hspace{3cm}\\
%\vskip.35in   
 $^{1}$~{\small Univ. di Padova and INFN-Padova, Via Marzolo 8, 
I-35131 Padua, Italy}.
\hspace{0.3cm}\\
 $^{2}$~{\small I.E.M. (CSIC), Serrano 123, 28006 Madrid, Spain}.
\hspace{0.3cm}\\
 $^{3}$~{\small I.F.T. C-XVI, U.A.M., 28049 Madrid, Spain}.
\hspace{0.3cm}\\
 $^{4}$~{\small I.M.A.F.F. (CSIC), Serrano 113 bis, 28006 Madrid, Spain}.
\hspace{0.3cm}\\
 $^{5}$~{\small IPPP, University of Durham, DH1 3LE Durham, UK}.
} 
\date{} 
\maketitle 
\def\baselinestretch{1.15} 
\begin{abstract}
\noindent
We consider supersymmetric scenarios in which the scale of SUSY 
breaking is low, $\sqrt{F}={\cal O}(\rm TeV)$. Instead of
studying specific models of this type, {\it e.g.}~those with extra
dimensions and low fundamental scale, we follow a 
model-independent approach based on a general effective 
Lagrangian, in which the MSSM supermultiplets are effectively
coupled to a singlet associated to SUSY breaking. 
Our goal is to analyse the interplay bewteen SUSY 
breaking and electroweak breaking, generalizing earlier
results. The conventional MSSM picture can be substantially
modified, mainly because the Higgs potential contains additional
effective quartic terms and resembles that of two-Higgs-doublet 
models, with an additional singlet. Novel opportunities to achieve 
electroweak breaking arise, and the electroweak scale may be 
obtained in a less fine-tuned way. 
Also the Higgs spectrum can be strikingly changed, and the lightest 
state can be much heavier than in usual supersymmetric 
scenarios. Other effects appear in the chargino and neutralino
sectors, which contain the goldstino. Finally, we discuss
the role of electroweak breaking in processes in which
two goldstinos could be emitted, such as 
fermion-antifermion annihilation and the invisible decay 
of a $Z$ boson or of neutral Higgs bosons.
\end{abstract}

\thispagestyle{empty}
\vspace*{0.1cm}
\leftline{January 2003}
\leftline{}

\vskip-24cm
\rightline{DFPD-02/TH/34}
\rightline{IFT-UAM/CSIC-02-60}
\rightline{IEM-FT-229/02}
\rightline{IPPP/02/84}
\rightline{DCPT/02/168}
\rightline{hep-ph/0301121}
\vskip3in

\end{titlepage}
%%%%%%%%%%%%%%%%%%%%%%%%%%%%%%%%%%%%%%%%%%%%%%%%%%%%%%%%%%%%%%%%%%%
\setcounter{footnote}{0} \setcounter{page}{1}
\newpage
\baselineskip=20pt

\noindent

\section{Introduction}
\setcounter{equation}{0}
\renewcommand{\theequation}{1.\arabic{equation}}

The Minimal Supersymmetric Standard Model (MSSM) \cite{MSSM} has
been for many years the paradigm of phenomenologically viable
supersymmetric (SUSY) theories. In the MSSM the observable matter 
content is minimal and the breaking of SUSY (\SUSY) takes place 
in another sector of the fundamental theory and is then transmitted 
to the observable sector through some mediation mechanism.
As a result, one obtains a low-energy supersymmetric theory 
in which the MSSM multiplets are effectively coupled to 
the goldstino multiplet through non-renormalizable
interactions\footnote{In a model with minimal particle content
(besides the goldstino multiplet), if those effective interactions were
only of renormalizable type, the property ${\rm STr}{\cal M}^2=0$
would hold and the spectrum would not be realistic.}, {\it i.e.}~the 
effective K\"ahler potential and gauge kinetic function
are generically non-minimal. 
Well known examples of mediation mechanisms are supergravity mediation 
and gauge mediation, where such effective interactions
arise at the tree-level and at the loop-level respectively.  

In the usual MSSM, once SUSY is broken, the effective theory is
approximated by a renormalizable one, in which the \SUSY sector has
been decoupled, leaving as footprint a set of soft breaking terms that
arise from  the above effective interactions. In obtaining these soft
terms the superfields responsible for \SUSY play an external role,
through the expectation values of their auxiliary fields. The
approximation that the soft terms encode all the effects of \SUSY in
the observable sector is a good one when the scale of \SUSY mediation,
$M$, is very large.
However, in scenarios where $M$ is low 
(not far from the TeV scale) this picture might be
not accurate enough, and the `hidden sector' might be not so hidden.
This already happens to some extent in gauge-mediated scenarios
(see {\it e.g.} \cite{Giudice:1998bp}), 
where $M\sim{\cal O}(10-10^3)$ TeV, and also, more
characteristically, in scenarios of extra dimensions (more or less
string-motivated) in which the fundamental scale is quite low, 
typically ${\cal O}$(TeV) (see {\it e.g.}~\cite{warped}). 
More generally, deviations from the conventional MSSM picture
appear whenever the low-energy supersymmetric effective theory
is obtained by integrating out physics at energy scales
not far from the TeV scale.
Let us briefly summarize how this comes about. 

In specific models, spontaneous \SUSY takes place in a sector where  the
auxiliary components of a set of fields get non-vanishing  
vacuum expectation values (VEVs).  In
the simplest cases this sector can be parametrized by a single chiral
superfield $T$.  Then, the effective interactions between $T$ and the
MSSM superfields produce at the same time: (i) SUSY breaking effects
among the MSSM  multiplets, as a consequence of the non-vanishing
$\langle F^T \rangle$;  
(ii) specific interactions between the MSSM multiplets and the physical
degrees of freedom in the $T$ multiplet,  {\it i.e.}~the goldstino
and its scalar partners (see {\it e.g.} \cite{Brignole:1996fn}).  
The form and size
of these effects depend crucially on the relation between the
mediation scale ($M$), the SUSY breaking scale ($\sqrt{F}$)  and the
electroweak scale ($M_W$), taking into account that the size of
induced \SUSY masses, $\tilde{m} \sim F/M$, should be ${\cal
O}$(TeV).  In the case of a strong  hierarchy $M \gg \sqrt{F} \gg
M_W$, type (i) effects reduce to the so called `soft  breaking 
terms' \cite{soft} and type (ii) effects are  negligible.
This limit corresponds to the conventional MSSM.  
However, in the opposite case of mild (or no) hierarchy, 
{\it i.e.}~when $M$ and $\sqrt{F}$ are in the 
TeV range, novel type (i) effects emerge, such as the so-called 
`non-standard soft terms' and `hard breaking terms' 
\cite{hard,Polonsky}. These include, in particular,
${\cal O}(F^2/M^4)$ contributions to quartic Higgs couplings,
whose phenomenological impact was recently emphasized 
\cite{Polonsky}.
Moreover, type (ii) effects are no longer negligible and 
have important phenomenological consequences. 
For instance, goldstinos (or, equivalently \cite{Fayet:vd,cddfg}, 
light gravitinos) can appear in the decays of MSSM superparticles
already for moderate values of $\sqrt{F}$ \cite{hgh}.
For sufficiently low values of $\sqrt{F}$, goldstinos can also be 
directly produced in high energy collisions at non-negligible rates,
either in association with MSSM superparticles \cite{fay,dnw,dnwlnz} 
or even without them \cite{gcoll,Luty:1998np}. The scalar partners 
of the goldstino (sgoldstinos) can be produced as well 
\cite{brdrnw,dnw,sgol}.
Moreover, goldstinos and sgoldstinos can contribute to
$(g-2)_{\mu}$ \cite{gmu} and to several flavour changing
or flavour conserving transitions \cite{Brignole:2000wd,gorb},
and can also play a role in astrophysics and
cosmology (see {\it e.g.} \cite{astr}).

The main purpose of this paper is to further explore  
scenarios with low SUSY breaking scale and, in 
particular, to analyse the interplay between SUSY breaking
and electroweak breaking and to examine the Higgs sector. 
Since we work at the effective
theory level, the field content is very economical: the only addition
to the supersymmetrized Standard Model is a singlet field $T$
(responsible for  \SUSY), coupled non-minimally to the MSSM
superfields.  In section~2 we  recall some general aspects of the
effective description of SUSY breaking and make more explicit some of the
arguments presented above. After recovering standard formulae for the MSSM
mass parameters, we mention some effects expected in
non-hierarchical scenarios.  In section~3 we focus on the Higgs
sector, analyse the pattern of electroweak symmetry 
breaking showing that new options are possible, and 
discuss the general effective interactions between the Higgs
superfields and the $T$ superfield.  In particular, we study how SUSY
breaking effects can transform the conventional MSSM Higgs sector into a
less constrained one, closer to that of generic two-Higgs-doublet models. 
In section~4 we make a convenient choice of field coordinates
and give further details on the Higgs potential and the 
neutralino/chargino sectors. In section~5 we give, for illustrative 
purposes, two simple
examples of models with low \SUSY scale that have a small number of 
parameters. In section~6 we discuss
the effective interactions involving two goldstinos and SM particles,
and study how electroweak breaking affects such couplings.
Finally, we summarize our results in section~7.
In Appendix A we discuss the minimization of symmetric 
two-Higgs-doublet potentials.

\section{Effective supersymmetry breaking}
\setcounter{equation}{0}
\renewcommand{\theequation}{2.\arabic{equation}}

Throughout this paper we will describe SUSY breaking effects 
using an effective Lagrangian description, without 
relying on a specific microscopic dynamics,
in an approach analogous to \cite{Brignole:1996fn}.
More specifically we will assume that, after integrating out 
some fundamental degrees of freedom, we are left with an 
effective globally supersymmetric four-dimensional theory
whose degrees of freedom are the MSSM ones and a 
singlet chiral superfield associated with \SUSY.
Before discussing this, however, it is useful to 
recall some general properties of SUSY effective
Lagrangians. 

\subsection{General effective Lagrangian}

Let us consider a general $N=1$ globally supersymmetric theory 
in four dimensions, with gauge group $G$, vector superfields 
$V=V^a {\bf t}_a$ 
and chiral superfields $\phi^i$ (see {\it e.g.} \cite{wbggrs}).
The effective Lagrangian for such a theory has the 
general form
\be
\label{leffgen}
{\cal L} = 
\int \!\! d^4 \theta \left[ K (\bar{\phi}, e^{2 V} \phi) 
+ 2 \xi_a V^a \right]
+ \left[ \int \!\! d^2 \theta \, W(\phi) +  {\rm h.c.} \right]
+ {1 \over 4} \left[ \int \!\! d^2 \theta f_{ab}(\phi)
{\cal W}^a {\cal W}^b +  {\rm h.c.} \right] .
\ee
where $K(\bar{\phi},\phi)$, $W(\phi)$ and $f_{ab}(\phi)$ are the 
effective K\"ahler potential, superpotential and gauge kinetic 
functions, respectively, and higher derivative terms are neglected.
The Fayet-Iliopoulos parameters $\xi_a$ can be non-vanishing
for the abelian factors of $G$ and are shown here only for 
completeness (we will assume that the $\xi_Y$ vanishes).
The effective Lagrangian for the component 
fields\footnote{We decompose chiral superfields according to
$\phi^i \Rightarrow \phi^i + \sqrt{2} \psi^i \theta +  F^i \theta\theta 
+ \ldots$ and vector superfields
according to $V^a  \Rightarrow A^a_{\mu} 
\theta \sigma^{\mu} \bar{\theta} + 
(\lambda^a \theta \bar{\theta}\bar{\theta} + {\rm h.c.})
+ {1 \over 2} D^a \theta \theta \bar{\theta}\bar{\theta}$,
in the Wess-Zumino gauge. Notice that
we directly define as $\lambda^a$ what is often
introduced as $-i \lambda^a$ and then redefined.
Our space-time metric has signature $(+---)$
and we use two-component spinor notation, with
$\sigma^{\mu}=(1,\sigma^A)$,
$\bar{\sigma}^{\mu}=(1,-\sigma^A)$, where
$\sigma^A$ are the Pauli matrices.} can be
obtained by a standard procedure \cite{wbggrs}.
In particular, the scalar potential has the general expression
\be
\label{Vgeneral}
V=V_F + V_D = W_i K^{i\bar{\jmath}} \bar{W}_{\bar{\jmath}}
\ + \ 
{1\over 2} \left[K_i ({\bf t}_a\phi)^i + \xi_a \right]
f_R^{ab}
\left[K_j ({\bf t}_b\phi)^j + \xi_b \right] \, .
\ee
Subscripts denote derivatives ($W_i \equiv \partial W/\partial \phi^i$,
$\bar{W}_{\bar{\jmath}} \equiv \partial {\bar W}/
\partial \bar{\phi}^{\bar{\jmath}} \equiv (\partial W/\partial \phi^j)^*$,
$K_i \equiv \partial K/\partial \phi^i$,\ldots), 
$K^{i\bar{\jmath}}$ is the inverse of the K\"ahler metric 
$K_{\bar{\imath}j} \equiv \partial^2 
K/\partial \bar{\phi}^{\bar{\imath}} \partial \phi^j$ 
and $f_R^{ab}$ is the inverse of the metric $(f_R)_{ab} 
\equiv {\rm Re} f_{ab}$  of the vector sector
({\it i.e.} $K^{i\bar{\jmath}}K_{\bar{\jmath}\ell} =\delta^i_{\ell}$ 
and $f_R^{ab} (f_R)_{bc}=\delta^a_c$).
The order parameter for supersymmetry breaking, which will
be non-zero by assumption, is
\be
\label{fdef}
F^2 \equiv \langle V \rangle =  \langle V_F + V_D \rangle 
= \langle {\bar F}^{\bar{\imath}} K_{\bar{\imath}j} F^j
+ {1 \over 2} D^a (f_R)_{ab} D^b \rangle \, ,
\ee
where the VEVs of the auxiliary fields are
\be
\label{auxvevs}
\langle F^i \rangle = 
- \langle K^{i\bar{\jmath}}\bar{W}_{\bar{\jmath}} \rangle
 \;\; , \;\;
\langle D^a \rangle =
- \langle f_R^{ab} \left[K_j ({\bf t}_b\phi)^j + \xi_b \right] \rangle
\ .
\ee
We also recall that fermion mass terms have the form 
$ - {1 \over 2} (\lambda^a,\psi^i) {\cal M}  (\lambda^b,\psi^j)^T 
+ {\rm h.c.}$, where the matrix ${\cal M}$ is given by 
\be
\label{generalM}
\hspace{-0.5cm} {\cal M} = \left(
\begin{array}{cc}
- \displaystyle{1\over 2} \langle (f_{ab})_{\ell} F^{\ell} \rangle
&  
\sqrt{2} \langle K_{\bar{\ell} j} (\overline{{\bf t}_a \phi})^{\bar{\ell}}
+ \displaystyle{1\over 4}  (f_{ac})_j D^c \rangle
\\ &  \\
 \sqrt{2} \langle K_{\bar{\ell} i} (\overline{{\bf t}_b \phi})^{\bar{\ell}}
+ \displaystyle{1\over 4}  (f_{bc})_i D^c \rangle
&
\langle W_{ij} +  {\bar F}^{\bar{\ell}} K_{\bar{\ell}ij} \rangle
\end{array}
\right)\ . 
\ee
In particular, by using the extremum conditions of the scalar potential
and gauge invariance, it is easy to check that the mass matrix 
$ {\cal M}$ has an eigenvector 
$( {1\over \sqrt{2}} \langle D^b \rangle , \langle F^j \rangle)^T$
with zero eigenvalue, which corresponds to the goldstino 
state. This eigenvector specifies the components of
goldstino field $\tilde{G}$ contained in the original 
fields  $\psi^i$ and  $\lambda^a$, {\it i.e.} we have
\be
\label{gfrag}
\psi^i =  {\langle F^i \rangle \over F} \tilde{G} + \ldots
\;\; , \;\;
\lambda^a = {\langle D^a \rangle \over \sqrt{2} F} \tilde{G} 
+ \ldots \;\; ,
\ee
where the ellipses stand for other mass eigenstates\footnote{
The field $\tilde{G}$ appearing here is canonically 
normalized, whereas in general the fields $\psi^i$ and $\lambda^a$ 
are not. 
However, eq.~(\ref{gfrag}) can obviously be written in the same form
also in terms of canonically normalized fermion {\em and} auxiliary fields,
which are related to the original ones by the same rescaling:
$\langle K_{\bar{\imath}j} \rangle 
(i \bar{\psi}^{\bar{\imath}} \bar{\sigma}^{\mu} 
\partial_\mu \psi^j +  {\bar F}^{\bar{\imath}} F^j)
\rightarrow (i\bar{\psi}^{\bar{\imath}} \bar{\sigma}^{\mu} 
\partial_\mu \psi^i +  {\bar F}^{\bar{\imath}} F^i)$,
$ \langle (f_R)_{ab} \rangle (i\bar{\lambda}^a \bar{\sigma}^{\mu} 
\partial_\mu \lambda^b + {1 \over 2} D^a D^b)
\rightarrow (i \bar{\lambda}^a \bar{\sigma}^{\mu} 
\partial_\mu \lambda^a + {1 \over 2} D^a D^a)$.
Taking this step 
into account, one can invert eq.~(\ref{gfrag}) and express 
$\tilde{G}$ in terms of canonically normalized fields.}.
We also recall that, in the framework of local SUSY,
the goldstino degrees of freedom become the longitudinal
components of the gravitino, which obtains a mass
$m_{3/2}=F/(\sqrt{3}M_P)$, where $M_P$ is the Planck scale. 
When $\sqrt{F}$ is close to the electroweak scale, 
$m_{3/2}$ is much smaller than typical experimental energies, 
which implies that the dominant gravitino components are 
precisely the goldstino ones \cite{Fayet:vd,cddfg}.
 
%%%%%%%%%%%%%%%%%%%%%%%%%%%%%%%%%%%%%%%%%%%%%%%%%%%%%%%%%%%%%%

\subsection{High and low supersymmetry breaking scales}

In order to make contact with the usual MSSM framework, 
let us assume that the effective SUSY theory has 
gauge group $SU(3)\times SU(2) \times U(1)_Y$
and chiral superfields $\phi^i=(\phi^\alpha,T)$,
where $\phi^{\alpha}$ are the MSSM chiral
superfields (containing Higgses, leptons
and quarks) and $T$ is a singlet superfield 
whose auxiliary field VEV $\langle F^T \rangle$ 
breaks SUSY. For small fluctuations of the 
fields $\phi^{\alpha}$, the expansions of
$K$, $W$ and $f_{ab}$ read
\bea
\label{keff}
K & = & k(\bar{T},T) 
+ c_{\bar{\alpha} \beta} (\bar{T},T) \bar{\phi}^{\bar{\alpha}}\phi^\beta
+ {1 \over 2} \left[ d_{\alpha\beta}(\bar{T},T) \phi^\alpha \phi^\beta
+ {\rm h.c.}\right]
+ \ldots
\\
\label{weff}
W & = &  w(T) + {1\over 2} \mu_{\alpha\beta}(T) 
\phi^\alpha \phi^\beta  + {1 \over 3!} h_{\alpha\beta\gamma}(T) 
\phi^\alpha \phi^\beta \phi^\gamma + \ldots
\\
\label{feff}
f_{ab} & = & f_a(T) \delta_{ab} + \ldots
\eea
In the (zero-th order) vacuum defined by $\langle \phi^{\alpha} 
\rangle_0=0$, we have $F^2 \simeq  \langle V \rangle_0$
$= \langle |F^T|^2 k_{\bar{T} T} \rangle_0
= \langle |w_T|^2/ k_{\bar{T} T} \rangle_0$, which
by assumption is non-zero.
The functions $c_{\bar{\alpha} \beta}, d_{\alpha\beta}, 
h_{\alpha\beta\gamma}, f_a$ are assumed to
depend on $T$ through the ratio $T/M$, where 
$M$ is some (not yet determined) scale. Then the induced 
SUSY-breaking mass splittings within the $\phi^{\alpha}$ 
and $V^a$ multiplets are characterized by a scale 
$\tilde{m}\sim F/M$. We also make the standard assumption that
$\mu_{\alpha\beta}$, if non-vanishing, has size ${\cal O}(\tilde{m})$ 
rather than ${\cal O}(M)$, {\it i.e.}~$\mu_{\alpha\beta}(T) \sim (F/M) 
{\tilde\mu}_{\alpha\beta}(T/M)$.

If we fix $\tilde{m}$ to be $\cal{O}({\rm TeV})$, there is still
much freedom in choosing $M$ and $F$. 
Standard scenarios are characterized by a strong hierarchy 
$M \gg \sqrt{F} \gg \tilde{m}$.
In this limit the physical components of the $T$ multiplet
({\it i.e.}~the goldstino and its scalar partners, the  `sgoldstinos')
are almost decoupled from the other fields, and the effective
theory for the $\phi^{\alpha}$ and $V^a$ multiplets is well 
approximated by a renormalizable one. The latter is 
characterized by gauge couplings $g_a^2 = 1/\langle {\rm Re}f_a  
\rangle_0$, an effective superpotential $\hat W$ and a set of 
soft SUSY breaking terms, whose mass parameters are 
${\cal O}(\tilde{m})$. This is the usual MSSM scenario \cite{MSSM}.
The MSSM parameters can be computed in terms of the functions 
appearing in $K$, $W$ and $f_{ab}$ above. Let us 
consider for simplicity the case of diagonal matter metric,
{\it i.e.}~$c_{\bar{\alpha} \beta} = c_{\alpha} 
\delta_{\bar{\alpha} \beta}$, and 
rescale the fields in order to have canonical normalization:
$\langle \sqrt{c_\alpha} \rangle_0 \phi^\alpha \rightarrow \phi^\alpha$,
$\langle \sqrt{ {\rm Re}f_a } \rangle_0 V^a  \rightarrow V^a$. 
The effective superpotential of the renormalizable theory is
\be
\hat{W} = {1\over 2} \hat{\mu}_{\alpha\beta} \phi^\alpha \phi^\beta  
+ {1 \over 3!} \hat{h}_{\alpha\beta\gamma} \phi^\alpha \phi^\beta
\phi^\gamma\ ,
\ee
where 
\bea
\label{mupar}
\hat{\mu}_{\alpha\beta} & = &  \left\langle 
\frac{ \mu_{\alpha\beta} + \bar{F}^{\bar{T}} \partial_{\bar{T}} 
d_{\alpha\beta} }
{ (c_\alpha c_\beta)^{1/2} } \right\rangle_0 \ ,
\\
\hat{h}_{\alpha\beta\gamma} & = &  \left\langle 
\frac{ h_{\alpha\beta\gamma} }{ (c_\alpha c_\beta c_\gamma)^{1/2} } 
\right\rangle_0 \ .
\eea
Soft breaking terms are described by
\be
{\cal L}_{\rm soft} =  - \tilde{m}_\alpha^2 |\phi^\alpha|^2
- \left[ {1\over 2} (\hat{\mu}_{\alpha\beta} B_{\alpha\beta}) 
\phi^\alpha \phi^\beta 
+ {1 \over 3!} (\hat{h}_{\alpha\beta\gamma} A_{\alpha\beta\gamma}) 
\phi^\alpha \phi^\beta \phi^\gamma + 
 {1\over 2} M_a \lambda^a \lambda^a + {\rm h.c.} \right],
\ee
where 
\bea
\tilde{m}_\alpha^2 & = &  \left\langle - |F^T|^2 
\partial_T \partial_{\bar{T}}
\log c_\alpha \right\rangle_0 \ ,
\\
B_{\alpha\beta} & = & \left\langle - F^T  \partial_T 
\log \left( \frac { \mu_{\alpha\beta} + \bar{F}^{\bar{T}} 
\partial_{\bar{T}} d_{\alpha\beta} }
{c_\alpha c_\beta} \right) \right\rangle_0 \ ,
\\
A_{\alpha\beta\gamma} & = & \left\langle - F^T  \partial_T 
\log \left( \frac
{ h_{\alpha\beta\gamma} } {c_\alpha c_\beta c_\gamma} \right) 
\right\rangle_0 \ ,
\\
\label{mgaugi}
M_a & = & \left\langle - F^T  \partial_T \log ( {\rm Re} f_a) 
\right\rangle_0 \ .
\eea
The above results agree with \cite{Brignole:1996fn} and are
compatible with a specific limit ($M_P \rightarrow \infty$,
$m_{3/2} \rightarrow 0$ with $F= \sqrt{3}m_{3/2} M_P$ fixed)
of supergravity results \cite{SUGRA}.

Notice that the $T$ multiplet has played an external role
in the previous derivation: it has only provided the SUSY breaking 
VEV $\langle F^T \rangle$. Moreover, only the leading terms
in an expansion in $F/M^2$ have been retained, because of the
assumed hierarchy. Other terms are strongly suppressed.
However, if the scales $M$ and $F$ are not much larger than the 
TeV scale and the ratio $F/M^2 \sim \tilde{m}/M \sim \tilde{m}^2/F$ 
is not negligible, the standard MSSM picture is corrected by additional 
effects and novel features emerge. For instance, the components of MSSM
multiplets ($\phi^\alpha$ and $V^a$) can have novel non-negligible 
interactions among themselves as well as non-negligible interactions 
with the physical components of $T$ (goldstino and sgoldstinos),
as we have already recalled in the introduction.
Moreover, since some of the $\phi^\alpha$ fields ({\it i.e.}~the Higgses) 
have to obtain a VEV in order to break the gauge symmetry, 
in principle one should reconsider the minimization of 
the scalar potential taking into account both $T$ and such
fields. In addition, the $F$ components of the Higgs multiplets and
the $D$ components of the neutral vector multiplets
could give non-negligible contributions to SUSY breaking.
In this case the goldstino could have components along
all neutral fermions ($\tilde T$, Higgsinos and gauginos).
One could even conceive extreme scenarios in which the $T$ 
field is absent and the Higgs fields alone are effectively 
responsible for breaking both SUSY and the gauge symmetry 
(see {\it e.g.}~\cite{Brignole:1995cq,Brignole:1996fn}, where examples 
of this type with singular superpotentials were given). 
In this paper we consider scenarios in which both $T$ 
and the Higgs fields are present and show how unconventional 
features emerge.

\section{The Higgs sector}
\setcounter{equation}{0}
\renewcommand{\theequation}{3.\arabic{equation}}

We now focus on the MSSM Higgs sector, made
of two $SU(2)$ doublets $(H_1, H_2)$. When $F/M^2$ is
not negligible, some higher order terms in the expansions
of $K$, $W$ and $f_{ab}$ not 
written explicitly in eqs.~(\ref{keff}), (\ref{weff}) and 
(\ref{feff}) can become important. We will explicitly write
all the ${\cal O}(H^4)$ terms in $W$ and $K$, which
will be sufficient for our purposes. 
As anticipated in the previous section, the coefficient functions 
appearing in $W$ and $K$ will depend on the field $T$ and on some 
mass scales. 
Thus we write\footnote{The symbol $\cdot$ stands for the $SU(2)$ 
product: $H_1\cdot H_2 = H_1^0 H_2^0 - H_1^- H_2^+$.}:
\bea
W & = & w(T) + \mu(T) H_1\cdot H_2 + {1 \over 2 M}\ell(T)\left(H_1
\cdot H_2\right)^2 
+ \ldots
\label{W}
\\
K & = & 
k(\bar{T},T) + c_1(\bar{T},T)|H_1|^2  + c_2(\bar{T},T)|H_2|^2 
+ \left[ d(\bar{T},T) H_1\cdot H_2 + {\rm h.c.}\right]
\nonumber
\\
& + & {1\over 2M^2} e_1(\bar{T},T)|H_1|^4 
+{1\over 2M^2} e_2(\bar{T},T)|H_2|^4
+{1\over M^2}e_3(\bar{T},T)|H_1|^2 |H_2|^2 
\nonumber
\\
& + & {1\over M^2}e_4(\bar{T},T)|H_1\cdot H_2|^2 
+\left[ {1 \over 2M^2} e_5(\bar{T},T) (H_1\cdot H_2)^2\right.
\nonumber 
\\ 
&+&\left. {1\over M^2}e_6(\bar{T},T)|H_1|^2 H_1\cdot H_2
+{1\over M^2}e_7(\bar{T},T)|H_2|^2 H_1\cdot H_2 
+  {\rm h.c.} \right]+ \ldots
\label{Kahler}
\eea
The K\"ahler potential $K$ is assumed to contain a 
single mass scale $M$. Thus the coefficient functions 
$c_i$, $d$ and $e_i$ in $K$ are in fact dimensionless functions of 
$T/M$ and $\bar{T}/M$ while
$k(\bar{T},T) \sim M^2 \tilde{k} (\bar{T}/M,T/M)$. 
On the other hand, $W$ should contain, besides $M$, the SUSY-breaking
scale $F$ (notice that $F\sim \langle \partial_T W \rangle$). 
Although it is
not possible to determine from first principles what is the precise
dependence on $M$ and $F$ of the coefficient functions in $W$, a
reasonable criterion is to insure that each parameter of the component
Lagrangian in the $T$-$H_1$-$H_2$ sector receives contributions 
of the same order from  $K$ and
$W$. An example of this are the two contributions to the effective
$\hat{\mu}$ parameter in Eq.~(\ref{mupar}).  The plausibility of this
criterion is  stressed by the fact that there is a considerable
freedom to move terms between $K$ and $W$ through analytical
redefinitions of the superfields (see subsection 4.1 below).  
Consequently, we can assume\footnote{This digression refers 
to generic values of the ratio $F/M^2$, in the spirit of the 
general discussion presented so far. It is clear that the assumed 
scale dependences are more meaningful 
for $F/M^2 \ll 1$ than for $F/M^2 \sim {\cal O}(1)$. However,
having a scaling rule for generic scenarios can be 
a useful book-keeping device ({\it e.g.} to interpolate different
cases).}
\be
\label{wscal}
w(T) \sim F M \tilde{w}(T/M) \ , \;\;
\mu(T) \sim {F \over M} \tilde{\mu}(T/M) \ , \;\;
\ell(T) \sim {F \over M^2} \tilde{\ell}(T/M) \ ,
\ee
where $ \tilde{w},\tilde{\mu},\tilde{\ell}$ are
dimensionless functions of their arguments.
The above dependences can be motivated by a broken $U(1)_R$ 
symmetry under which the fields $T,H_1,H_2$ 
and the parameter $M$ have zero charge, while $F$ has
$R$-charge 2 and acts as breaking parameter. Also notice 
that any ($M$-dependent) non-linear field redefinition of 
$T,H_1,H_2$ obviously respects this charge assignement.

For the expansion of the gauge kinetic functions $f_{ab}$ it 
is enough for our purpose to keep ${\cal O}(H^2)$ terms. 
Before writing this expansion, we recall that the indices
in $f_{ab}$ are saturated with those of the super-field-strengths 
${\cal W}^a {\cal W}^b$, see eq.~(\ref{leffgen}).
Thus the allowed irreducible representations in $f_{ab}$ are 
those contained in the symmetric product of two adjoints.
For the $SU(2)\times U(1)$ gauge group, such representations
are singlet, triplet and fiveplet:
\be
f_{ab} = f^{(s)}_{ab}+ f^{(t)}_{ab}+ f^{(f)}_{ab}
\ee
All these parts can be present, once $f_{ab}$ is allowed 
to depend on $H_1$ and $H_2$. 
The expansion of the singlet part $f^{(s)}_{ab}$ 
reads\footnote{
The singlet part is also present for the colour 
group, of course.}
\be
\label{fsin}
f^{(s)}_{ab}=\delta_{ab}\left[f_a(T)+{1 \over M^2} 
h_a(T) \, H_1\cdot H_2 + \ldots \right] \, .
\ee
The triplet part $f^{(t)}_{ab}$ is associated with
the $SU(2)$-$U(1)_Y$ cross-term ${\cal W}^A {\cal W}^Y$,
where $A$ is an $SU(2)$ index. Thus the non-vanishing
components of $f^{(t)}_{ab}$ are $f^{(t)}_{AY}=f^{(t)}_{YA}$
and their expansion starts at ${\cal O}(H^2)$: 
\be
\label{ftri}
f^{(t)}_{AY} = {1 \over M^2} \omega(T)\, ( H_1\cdot \sigma^A H_2 ) 
+ \ldots
\ee
In eqs.~(\ref{fsin},\ref{ftri}) we have 
inserted appropriate powers of $M^2$ as before,
and we can assume that $f_a$, $h_a$ and $\omega$ are 
dimensionless functions of $T/M$.
Finally, the fiveplet part $f^{(f)}_{ab}$ has
both indices in $SU(2)$. We will neglect this
part since its leading term is ${\cal O}(H^4)$.

\subsection{Scalar potential and electroweak breaking}

{}From $W$, $K$ and $f_{ab}$ one can compute the component Lagrangian,
and in particular the scalar potential, which is given 
by the general expression in eq.~(\ref{Vgeneral}).
It is clear that the expanded form of $V$ will be similar 
to that of $K$. More precisely, $V$ has the same form as 
in a two-Higgs-doublet model\footnote{For a recent
analysis of two-Higgs-doublet models, see {\it e.g.}
\cite{Gunion:2002zf} and references therein.} (2HDM), with 
$T$-dependent coefficients, {\it i.e.}
\bea
\label{VTH}
V & = & 
V_0(\bar{T},T) + m_1^2(\bar{T},T)|H_1|^2  + m_2^2(\bar{T},T)|H_2|^2 
+ \left[ m_3^2(\bar{T},T) H_1\cdot H_2 + {\rm h.c.}\right]
\nonumber
\\
& + &  
{1\over 2} \lambda_1(\bar{T},T)|H_1|^4 
+{1\over 2} \lambda_2(\bar{T},T)|H_2|^4
+\lambda_3(\bar{T},T)|H_1|^2 |H_2|^2 
+\lambda_4(\bar{T},T)|H_1\cdot H_2|^2 
\nonumber
\\
& + & 
\left[ {1 \over 2} \lambda_5(\bar{T},T) (H_1\cdot H_2)^2 
+\lambda_6(\bar{T},T)|H_1|^2 H_1\cdot H_2
+\lambda_7(\bar{T},T)|H_2|^2 H_1\cdot H_2 
+  {\rm h.c.} \right]\nonumber\\
&+&\ldots
\eea
where we have truncated at ${\cal O}(H^4)$. The parametric dependence 
of the coefficients in $V$ is $m_i^2 \sim {\cal O}(F^2/M^2)$ and 
$\lambda_i \sim  {\cal O}(F^2/M^4) +  {\cal O}(g^2) $.
Explicit expressions for $m_i^2(\bar{T},T)$ can be
deduced from the results of sect.~2.2, whilst
the form of the coefficients $\lambda_i(\bar{T},T)$ will
be discussed in detail in sect.~4.2. 

In general, for a given potential, one can try to 
perform either an exact minimization or at least an
iterative one, relying on the expansion of the
potential in powers of $H_i/M$ and on the consistent 
assumption that the Higgs VEVs are smaller than $M$, 
possibly through some tuning. In the iterative approach, 
the starting point for the determination of the VEVs are 
the zero-th order values 
of $\langle H_i^0 \rangle_0$ and $\langle T \rangle_0$,
where $\langle H_i^0 \rangle_0 =0$ and $\langle T \rangle_0$ 
is the minimum\footnote{We will assume that $V_0(\bar{T},T)$
determines $\langle T \rangle_0$ and gives 
${\cal O}(F^2/M^2)$ masses to the associated scalar 
fluctuations. This situation, which can be regarded as generic,
is also supported by naturalness considerations \cite{Brignole:1998uu}. 
We will not discuss the alternative possibility of constant 
$V_0(\bar{T},T)$. In this special case, $T$ would be a modulus 
at lowest order (even after the breaking of SUSY), and the
minimization of $V(T,H_1,H_2)$ should (or could)
simultaneously determine $\langle T \rangle$, 
$\langle H_1 \rangle$ and $\langle H_2 \rangle$. 
This interesting situation is more delicate from the viewpoint of an 
iterative solution, although it could be dealt with in specific 
models.} of $V_0(\bar{T},T)$.

In sections 4, 5 and 6 we will discuss in more detail the 
scalar potential, its minimization and other 
phenomenological implications, both in general and 
through specific examples. Here we note that 
the form of the Higgs potential in eq.~(\ref{VTH})
already allows us to make some general observations 
on the possible patterns of electroweak breaking.
Let us set $m_i^2\equiv \langle m_i^2(\bar{T},T) \rangle_0$ 
for brevity.
There are two necessary conditions for electroweak breaking, 
which for a polynomial Higgs potential imply the 
existence of a non-trivial minimum.

%%%%%%%%%%%%%%%%%%%%%%%%figure%%%%%%%%%%%%%%%%%%%%%%%%
\begin{figure}[t]
%\psdraft
\centerline{
\psfig{figure=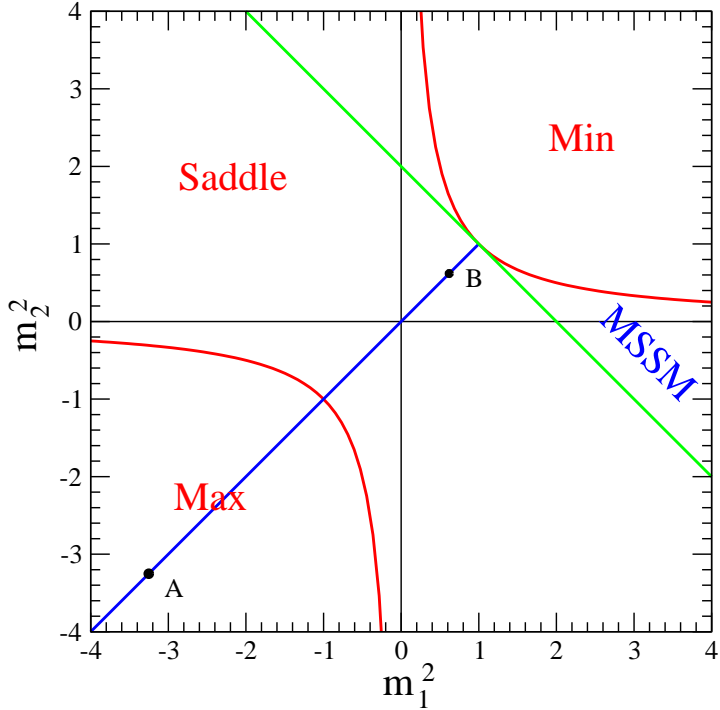,height=10cm,width=10cm,angle=-90,bbllx=3.cm,%
bblly=2.cm,bburx=20.cm,bbury=19.cm}}
\caption
{\footnotesize Schematic representation of the different possibilities for
electroweak breaking in a two-Higgs-doublet model, depending on the values
of $m_1^2$ and $m_2^2$ (with the axes in units of $|m_3^2|$). See text for
details.}
\label{thdmbreak}
\end{figure}
%%%%%%%%%%%%%%%%%%%%%%%%%figure%%%%%%%%%%%%%%%%%%%%%%%%

The first condition regards the origin of Higgs-field space.
This is a minimum, a saddle point or a maximum, 
depending on the mass parameters $m_i^2$:
\bea
m_1^2 m_2^2 - |m_3^2|^2 &>& 0\ ,\ m_1^2+ m_2^2 \ >\  0\ 
.\hspace{1.8cm} [\mathrm{Minimum}]
\label{minimum}
\\
m_1^2 m_2^2 - |m_3^2|^2 &<& 0\ , 
\hspace{4.2cm} [\mathrm{Saddle\  Point}]
\label{saddle}
\\  m_1^2 m_2^2 - |m_3^2|^2 &>& 0\ ,\ m_1^2+ m_2^2 \ <\  0\ 
.\hspace{1.8cm} [\mathrm{Maximum}]
\label{maximum}
\eea
These equations define three regions in the $\{m_1^2,m_2^2\}$-plane, 
labelled by  `Min', `Saddle' and `Max' in Fig.~1. Such regions
are separated by the upper and lower branches of the hyperbola 
$m_1^2 m_2^2 -|m_3^2|^2 =0$. Electroweak breaking 
can take place in the regions `Saddle' or `Max', while the region `Min'
is excluded\footnote{Actually, electroweak breaking could occur 
even in the case in which the origin is a minimum, 
through tunneling to a deeper non-trivial minimum. 
Models with such potentials have been considered in the literature, 
see {\it e.g.}~\cite{tunel}, but we will not discuss this
possibility.}.

The second condition for proper electroweak breaking
is the absence of unbounded from below directions (UFB) along which 
the quartic part of the Higgs potential gets destabilized. 
As a matter of fact the complete Higgs potential is necessarily bounded
from below since the full supersymmetric potential (\ref{Vgeneral}) is
positive definite. However, this does not guarantee that the 
truncation of $V$ at ${\cal O}(H^4)$, {\it i.e.}~eq.~(\ref{VTH}), 
is positive as well.  If it is not, this means that the positivity 
of the potential is ensured by higher order terms and the minima 
correspond to large values of $H_i$, which is not phenomenologically 
acceptable. 
UFB directions of this kind are normally prevented by quartic couplings.
In the MSSM the latter receive only  contributions from
D-terms, namely  $\lambda_{1,2}={1\over 4}(g^2 + g_Y^2)$,
$\lambda_3={1\over 4}(g^2 - g_Y^2)$,  $\lambda_4=-{1\over 2}g^2$,
$\lambda_{5,6,7}=0$. Then the potential  is indeed stabilized by the
quartic terms, except along the D-flat directions
$|H_1|=|H_2|$. Consequently, it is required 
that the quadratic part of $V$ be
positive along these directions:
\bea m_1^2 + m_2^2 - 2 |m_3^2| > 0\ .\hspace{2cm} [\mathrm{Potential\
 bounded\ from\ below}]
\label{UFB}
\eea
This condition applies {\em only} to the MSSM and
corresponds to the region of Fig.~1 above the straight line tangent 
to the upper branch of the hyperbola. Since eq.~(\ref{UFB}) 
is incompatible  with eq.~(\ref{maximum}), 
it follows that the MSSM conditions for electroweak  breaking 
are given by eqs.(\ref{saddle},\ref{UFB}), as is well known.
In Fig.~1 the corresponding region is a subset of the region
`Saddle' and is labelled by `MSSM':
it is made of the (two) areas between the upper branch of
the hyperbola and the tangent line.

However, when SUSY is broken at a moderately low scale, 
the $\lambda_i$ couplings in (\ref{VTH}) can also
receive sizeable ${\cal O}(F^2/M^4)$ contributions,
besides the ${\cal O}(g^2)$ ones.
Therefore condition (\ref{UFB}) is no longer mandatory to 
avoid UFB directions, since the boundedness of the potential
can be ensured by imposing appropriate conditions
on the $\lambda_i$ parameters\footnote{
The requirement of unbroken electric charge also 
imposes constraints \cite{ecb}.}.  
Thus the presence of the latter parameters extends the parameter 
space, relaxes the constraints on the quadratic part of the potential 
and opens a lot of new possibilities for electroweak breaking. 
In particular, both alternatives
(\ref{saddle}, \ref{maximum}) are now possible.
This means that most of the $\{m_1^2, m_2^2\}$ plane can in 
principle be explored: the whole regions labelled by `Saddle' and `Max'
in Fig.~1 are allowed, only the region `Min' is excluded.
This has several important consequences, that differ 
from usual MSSM results, and which we list below.
\begin{description}

\item[\bf (a)] 
The universal case $m_1^2= m_2^2$ is now allowed, unlike in the
MSSM. Actually, in the MSSM these mass parameters could be
degenerate at high energy and reach non-degenerate values 
radiatively by RG running (falling in the region `MSSM' of Fig.~1, 
typically with $m_1^2>0$, $m_2^2<0$).  The
fact that $m_2^2$ is the only scalar mass that tends to get negative
in this process is considered one of the virtues of the MSSM, in the sense 
that $SU(2)\times U(1)$
breaking is  ``natural''. Now, we see that even if the universal
condition holds at low-energy we can still break  
$SU(2)\times U(1)$. 
This will be illustrated in section~5 with two examples,
which correspond to points A and B in Fig.~1. We will also show that
$m_1^2= m_2^2$ does not necessarily imply $\langle |H_1|\rangle =
\langle |H_2|\rangle$, {\it i.e.}~$|\tan\beta|=1$.

\item[\bf (b)] 
Electroweak breaking generically occurs already at
tree-level. Still,
it is ``natural'' in a sense similar to the MSSM.
For example, if all the scalar masses are positive and universal,
$SU(2)_L\times U(1)_Y$  is the only symmetry that can be broken because
(with R-parity conserved) the only off-diagonal bilinear coupling among
MSSM fields is $m_3^2 \, H_1\cdot H_2$, which can drive symmetry
breaking in the Higgs sector if condition (\ref{saddle}) is satisfied.
This is just an example: we stress again that many unconventional
possibilities for electroweak breaking are allowed, including 
those in which both $m_1^2$ and $m_2^2$ are negative
and $m_3^2$ plays a minor role.

\item[\bf (c)] 
Finally, the fact that quartic couplings are very different from those
of the MSSM changes dramatically the Higgs spectrum and properties
(which will be tested at colliders, see {\it e.g.} 
\cite{testexp,Carena:2002es}). In particular, as illustrated by 
later examples, the MSSM bound on the
lightest Higgs field does no longer apply. Likewise, the fact that these
couplings can be larger than the MSSM ones may reduce the amount
of tuning necessary to get the proper Higgs VEVs.
Concerning the latter property, suppose that $F^2/M^2$ is 
significantly larger than the phenomenologically required
value of $v^2$. In this case, as is well
known in the MSSM, only one combination of the $m_i^2$ is allowed 
to be as large as ${\cal O}(F^2/M^2)$, whereas two other 
combinations should be tuned to values ${\cal O}(\lambda_i v^2)$, 
as a consequence of the minimization conditions. 
The interesting point is that the ${\cal O}(F^2/M^4)$ contributions
to the couplings $\lambda_i$ can exceed the familiar ${\cal O}(g^2)$ 
contributions, so the amount of fine tuning can be somewhat 
alleviated.

\end{description}

\subsection{Derivative couplings and the $\rho$-parameter}

In addition to modifications in the Higgs potential, the
non-renormalizable terms in the K\"ahler potential (\ref{Kahler})
generate derivative couplings. 
Explicitly, the generalized kinetic Lagrangian for 
$H_1, H_2, T$ reads
\bea 
\label{lkine}
{\cal L}_{kin}&=&|D_\mu H_1|^2\left[
c_1+{e_1\over M^2}|H_1|^2+{e_3\over M^2}|H_2|^2+\left( {e_6\over M^2}
H_1\cdot H_2+{\mathrm h.c.}\right)\right] \nonumber\\ 
&+&|D_\mu H_2|^2\left[ c_2+{e_2\over M^2}|H_2|^2+{e_3\over 
M^2}|H_1|^2+\left(
{e_7\over M^2} H_1\cdot H_2+{\mathrm h.c.}\right)\right] \nonumber\\
&+&|\partial_\mu T|^2\left[k_{T\bar T}+({c_1})_{T\bar T} |H_1|^2
+ ({c_2})_{T\bar T} |H_2|^2
+\left(d_{T\bar T} H_1\cdot H_2+{\mathrm h.c.}\right)
\right]\nonumber\\
&+&{e_1\over M^2}|H_1^\dagger D_\mu H_1|^2+{e_2\over 
M^2}|H_2^\dagger D_\mu H_2|^2 \nonumber\\ &+&{e_3\over 
M^2}\left[H_1^\dagger D_\mu H_1(D^\mu {H}_2)^\dagger H_2
+{\mathrm h.c.}\right]+{e_4\over M^2}
|\partial_\mu (H_1\cdot H_2)|^2 \nonumber\\
&+&\left\{\left[{e_6\over M^2}(D^\mu {H}_1)^\dagger H_1
 +{e_7\over M^2} (D^\mu {H}_2)^\dagger H_2\right]
\partial_\mu (H_1\cdot H_2)+{\mathrm h.c.}\right\}\nonumber\\
&+&\left\{\partial_\mu T^*\left[
({c_1})_{\bar T} H_1^\dagger D^\mu H_1+
({c_2})_{\bar T} H_2^\dagger D^\mu H_2+
d_{\bar T} \partial^\mu (H_1\cdot H_2)\right]+
{\mathrm h.c.} \right\}
\nonumber\\
&+& \ldots \ ,
\eea 
where
$|X_\mu|^2\equiv X_\mu^* X^\mu$ and
$c_i$, $e_i$, $k$, $d$ are the $T,\bar{T}$-dependent functions 
that appear in (\ref{Kahler}). Thus the Higgses and the
$T$ scalars also have derivative interactions, besides the
non-derivative ones described by the scalar potential.
Moreover, since the derivatives are gauge-covariant,
non-renormalizable interactions between scalar and 
vector fields appear as well. 

One of the consequences of electroweak symmetry breaking 
[$\langle H_i^0 \rangle=v_i/\sqrt{2}$, $\tan\beta\equiv v_2/v_1$, 
$v^2\equiv v_1^2+v_2^2$] is that
${\cal L}_{kin}$ generates mass terms for the gauge bosons.
Let us normalize the Higgs fields so that $\langle c_i \rangle_0 = 1$,
which implies $\langle c_i \rangle = 1 + {\cal O}(v^2/M^2)$,
and assume real parameters for simplicity.
We also temporarily neglect the non-singlet parts of $f_{ab}$, 
so gauge couplings are defined by 
$\langle {\rm Re}f_{ab}^{(s)} \rangle = g_a^{-2} \delta_{ab}$.
The gauge boson masses are: 
\bea 
M_W^2& = &  {1\over 4}g^2{\hat v}^2\ ,\\
M_Z^2&=&{1\over 4}(g^2+g_Y^2)\left[{\hat v}^2+{v^4\over 2M^2} \langle
e_1 c_\beta^4+e_2 s_\beta^4- 2 e_3 s^2_\beta c^2_\beta \rangle + \ldots 
\right]\ , 
\eea 
with 
\bea {\hat v}^2& = & v^2 \left[ 
\langle  c_1 c^2_\beta + c_2 s^2_\beta \rangle
+{v^2\over2M^2}
\langle e_1 c_\beta^4 + e_2 s_\beta^4 +2 e_3  s^2_\beta c^2_\beta
+ ( e_6 c_\beta^2  + e_7 s^2_\beta ) s_{2\beta}\rangle +\ldots 
\right]
\nonumber\\ 
&=&v^2 \left[1 + {\cal O}(v^2/M^2)\right] \ . 
\eea 
We see first that there is a small deviation of
the Higgs VEV $v$ from ${\hat v}=246$ GeV [of relative
order ${\cal O}(v^2/M^2)$, which we will ignore in the following], 
and second there is a
non-zero contribution to $\Delta\rho$ ({\it i.e.} $\epsilon_1$
or $\alpha \Delta T$), given by 
\be
\label{drho}
\Delta\rho= -{v^2\over 2M^2}\left[ c_\beta^4 \langle e_1 \rangle_0 
+ s_\beta^4  \langle e_2 \rangle_0 
-2 s_\beta^2 c_\beta^2 \langle e_3 \rangle_0 \right] 
+{\cal O}(v^4/M^4)\ .  
\ee 
This combination of parameters is constrained to
be small by electroweak precision measurements
\cite{PDG}, which could be used, for instance, to infer
a lower bound on the scale $M$, for given values of 
$\langle e_i \rangle_0$ and $\tan\beta$, or to constrain
the parameters $\langle e_i \rangle_0$, 
for given $M$ and $\tan\beta$.
Notice that a natural suppression of $\Delta\rho$ is 
obtained if the K\"ahler potential 
has an approximate $SU(2)_L\times SU(2)_R$ symmetry, 
since the latter implies the equality $e_1 = e_2=e_3$,
and also $|\tan\beta|=1$ after electroweak breaking
(see also Appendix A). 

Another set of non-renormalizable interactions
between scalar and vector fields originate from the 
gauge kinetic terms, {\it i.e.} from the field dependence
of the kinetic functions $f_{ab}(T,H_1,H_2)$.
In particular, the $T$ dependence of  $f_{ab}$, 
which is responsible for the leading contribution
to gaugino masses [eq.~(\ref{mgaugi})] and
to goldstino-gaugino-gauge boson couplings,
also induces interactions of the $T$ scalar field
with two gauge field strengths. The latter interactions 
are relevant, for instance, in the production and decays of 
$T$ scalars at colliders \cite{brdrnw,dnw,sgol}. Similarly, 
the dependence of $f_{ab}$ on $H_1$ and $H_2$ could have
interesting implications for the production and decays of Higgs 
bosons. In the latter case, the relevant part of $f_{ab}$ is
probably the singlet part $f_{ab}^{(s)}$, since 
the non-singlet parts are more constrained or suppressed.
In particular, we recall that the triplet part 
$f_{ab}^{(t)}$ of $f_{ab}$, eq.~(\ref{ftri}),
produces a kinetic mixing between the $SU(2)_L$ and the $U(1)_Y$
field strengths once the Higgs fields take VEVs.
This effect modifies the expressions of the gauge boson
masses and couplings, and one obtains 
a contribution to the $\epsilon_3$ parameter 
(or $\alpha S$) proportional to $s_\beta c_\beta 
\langle \omega \rangle_0 v^2/M^2$, which is
therefore constrained to be small.
Finally, we note that contributions to the $\epsilon_2$ 
parameter (or $\alpha U$) are automatically more suppressed, 
since they arise from the fiveplet part $f_{ab}^{(f)}$
of $f_{ab}$ and are ${\cal O}(s_\beta^2 c_\beta^2 v^4/M^4)$.

\section{General results in normal coordinates}
\setcounter{equation}{0}
\renewcommand{\theequation}{4.\arabic{equation}}

\subsection{Coordinate choices}

Theories with different expressions for $K$, $W$ and $f_{ab}$ have the 
same physical content if they are related by analytic redefinitions of the
chiral superfields.  This well known property is already clear, 
for instance, in
the results  for the MSSM mass parameters presented in section~2: the
spectrum only depends on specific combinations of the original
parameters. The effective $\hat{\mu}$ parameter of eq.~(\ref{mupar})
is a well known  example: its two `components' (from $W$ and $K$) can
easily be moved into one another by a redefinition of the $T$ field
that involves the Higgs fields. For instance,
if $K \supset |T|^2 - \left({\beta_\mu \over M} \bar{T} H_1\cdot H_2
+ {\rm h.c.} \right)$ and $W \supset \Lambda_S^2 T$, the
redefinition $T= T' + {\beta_\mu \over M} H_1\cdot  H_2$
leads to $K \supset |T'|^2 - {|\beta_\mu|^2 \over M^2} |H_1\cdot H_2|^2$
and $W \supset \Lambda_S^2 T' + \beta_\mu {\Lambda_S^2 \over M} H_1\cdot 
H_2 $.
Either coordinate choice leads to the same effective $\hat{\mu}$ 
parameter.

Sometimes it is better to avoid such field redefinitions, in order 
to keep track of all the different `sources' of a specific effective 
parameter or coupling. At other times, it is
convenient to exploit such redefinitions  in order to reduce the
redundant set of parameters to a minimal set. For instance,
one can try to remove as many terms as possible from 
the superpotential and reduce it to a minimal one. In our case 
we could first shift $T$ so that its zero-th order 
VEV vanish, $\langle T \rangle_0 = 0$, and then redefine 
the whole $W(T,H_1,H_2)$ to be just the new $T$ 
field, {\it i.e.} $W(T,H_1,H_2)= \Lambda_S^2 T'$,
where $\Lambda_S^2 \sim F$ (SUSY breaking scale). 
An advantage of this coordinate choice is that all the parameters 
in the component Lagrangian  automatically have a simple 
dependence  on $F$ and $M$, {\it e.g.}~$\hat{\mu} \sim F/M$.  
An example with such a minimal superpotential will be described 
in section~5. 
Another possibility, orthogonal to the previous one,
is to remove as many terms as possible from $K$. In general, one 
can set to zero all the derivatives
$K_{\bar{I} J_1 J_2 \ldots J_n}$ ($n>1$) and their conjugates around
a given point \cite{grk} (K\"ahler {\em normal coordinates}).  
In our case we could first shift $T$ so that $\langle T \rangle_0=0$ 
and then use normal coordinates around the origin\footnote{
The elimination of $\bar{T} H_1 H_2$ from $K$ illustrated 
above is a simple example of such a procedure.}.
In the next subsections we will make this choice and 
present explicit results on the scalar potential and the 
fermionic spectrum. Of course, it is also possible to employ 
intermediate coordinate choices, or even to make no coordinate 
choice at all, with equivalent results.

\subsection{The scalar potential}

The form of $W$ and $K$  [see Eqs.(\ref{W}, \ref{Kahler})] expressed
in normal coordinates and expanded in the $H_i$ and $T$ fields reads
\bea 
\label{wnor}
W & = &  \Lambda_S^2 \left( T + {1\over 6 M^2}\rho_t T^3 +\ldots \right)
+  \left(\mu + \mu' {T\over M} + {1\over 2}\mu''
{T^2\over M^2} +\ldots \right)H_1 \cdot H_2 
\nonumber\\ & + & 
{1\over 2 M} \left(\ell+\ell'{T\over M} +\ldots \right)
\left(H_1\cdot H_2\right)^2 +\ldots \\ 
\label{knor}
K & = &
\left( |T|^2-{1\over 4}\alpha_t{|T|^4\over M^2} +\ldots  \right)
+|H_1|^2\left[1+\alpha_1{|T|^2\over M^2}+ {1\over 2M^3}(\alpha'_1 T^2
\bar{T}+\bar{\alpha}'_1 T \bar{T}^2)+\ldots \right] \nonumber\\ & + &
|H_2|^2\left[1+\alpha_2{|T|^2\over M^2}+ {1\over 2M^3}(\alpha'_2T^2
\bar{T}+\bar{\alpha}'_2 T \bar{T}^2)+\ldots \right] \nonumber\\ & + &
\left[H_1\cdot H_2 \left({1\over 2}\alpha_3 {\bar{T}^2\over M^2} +
{1\over 2M^3}\alpha_3'  \bar{T}^2 T +\ldots \right) + {\rm h.c.}\right]
\nonumber\\ & + & {1\over M^2}\left\{ {1\over 2} |H_1|^4 \left[e_1 +
{1\over M}( e'_1 T+\bar{e}'_1 \bar{T})+e''_1 {|T|^2 \over M^2}
+\ldots \right]\right.\nonumber\\ & + & {1\over 2} |H_2|^4 
\left[e_2 +  {1\over M}(e'_2 T+\bar{e}'_2 \bar{T})+e''_2 {|T|^2\over M^2}
+\ldots \right]
\nonumber\\ & + & |H_1|^2|H_2|^2 \left[e_3 +  {1\over M}(e'_3
T+\bar{e}'_3 \bar{T})+e''_3 {|T|^2\over M^2}+\ldots \right] \nonumber\\ 
& + & |H_1\cdot H_2|^2\left[e_4+{1\over M}(e'_4 T+\bar{e}'_4 \bar{T})
+e''_4 {|T|^2\over M^2}+\ldots \right] \nonumber\\ & + & \left[|H_1|^2
H_1\cdot H_2\left(e'_6 {\bar{T}\over M}+e''_6 {|T|^2\over  M^2}
+\ldots \right) + {\rm h.c.}\right] \nonumber\\ 
& + &\left.  \left[|H_2|^2 H_1\cdot H_2\left(e'_7 {\bar{T}\over M}
+e''_7 {|T|^2\over M^2} +\ldots \right) + {\rm h.c.}\right]\right\} 
+\ldots \eea 
It is not restrictive to take $\Lambda_S^2$ real and positive. 
The coefficients of real invariants in $K$ ({\it e.g.} $\alpha_t, 
\alpha_1, \alpha_2, \ldots$) are necessarily real.
For the sake of generality we allow other parameters to be 
complex, keeping in mind that they should be taken as real 
if $CP$ conservation is imposed. 
The terms explicitly shown above are sufficient to compute all the   
${\cal O}(T^2)$, ${\cal O}(H^2)$, ${\cal O}(T H^2)$,
and ${\cal O}(H^4)$ terms of the scalar potential,
which in turn are sufficient\footnote{For instance, in these coordinates 
$e_5(\bar{T},T) \sim e''_5 \bar{T}^2/M^2 + \ldots$ is not
shown because it gives higher order corrections. 
Concerning the lowest order terms in $W$, notice that $\Lambda_S^2 T$ 
is necessarily present in order to break SUSY. 
Once this is taken into account, the absence of the 
${\cal O}(T^2)$ term is just a consequence of our coordinate 
choice and of the zero-th order minimization conditions. 
Indeed, the condition $\langle T \rangle_0 = 0$ relates the 
coefficient of the ${\cal O}(T^2)$ term in $W$ to the coefficient 
of the ${\cal O}(T^2 \bar{T})$ term in $K$, which is zero
in normal coordinates.} to evaluate VEVs and spectrum 
at lowest non-trivial order in $v/M$. Also notice that,
to this purpose, it is sufficient to keep the zero-th order part 
of $f_{ab}$, {\it i.e.} $\delta_{ab}/g_a^2$. We postpone the explicit 
expansion of $f_{ab}$ to the discussion of fermion masses below.
It is also convenient to define the auxiliary quantity
\be 
\tilde{m}\equiv {\Lambda_S^2\over M}\ , 
\ee 
which 
controls the typical magnitude of SUSY-breaking masses and 
appears frequently in what follows. In doing 
parametric estimates we will also apply eq.~(\ref{wscal}),
{\it i.e.} we will implicitly assume $\mu,\mu',\mu'' ={\cal O}(\tilde{m})$ 
and $\ell,\ell' ={\cal O}(\tilde{m}/M)$.

The scalar potential $V=V_F + V_D$ can be computed
from the general expression (\ref{Vgeneral})
and expanded as in eq.~(\ref{VTH}). The latter 
expansion can be further specialized using the above 
parametrization of $W$ and $K$.
We obtain:
\bea
\label{Vnorcoord}
V & = & \Lambda_S^4 + \alpha_t \tilde{m}^2 |T|^2 
+  {1\over 2}(\rho_t  \tilde{m}^2 T^2  + {\rm h.c.})
+ m_1^2 |H_1|^2 +  m_2^2 |H_2|^2 +  \left(m_3^2
H_1\cdot H_2 + {\rm h.c.}\right) \nonumber\\ & + & (a_1 T +
\bar{a}_1\bar{T} ) |H_1|^2  + (a_2 T + \bar{a}_2\bar{T} ) |H_2|^2 +
\left[ (a_3 T + a_4 \bar{T} ) H_1\cdot H_2 + {\rm h.c.} \right]
\nonumber\\ & + & {1\over 2} \lambda_1 |H_1|^4 + {1\over 2} \lambda_2
|H_2|^4  +  \lambda_3 |H_1|^2 |H_2|^2 + \lambda_4 |H_1\cdot H_2|^2
\nonumber\\ & + & \left[ {1 \over 2} \lambda_5 \left(H_1\cdot
H_2\right)^2   + \lambda_6 |H_1|^2 H_1\cdot H_2  + \lambda_7 |H_2|^2
H_1\cdot H_2 + {\rm h.c.} \right]+\ldots \eea

The coefficients of the ${\cal O}(T^2)$ terms should satisfy 
$\alpha_t > |\rho_t|$, for consistency with the condition
$\langle T \rangle_0=0$. The two degrees of freedom of the complex
field $T$ have masses $m_{T_{1,2}}^2= (\alpha_t \pm |\rho_t|) 
\tilde{m}^2 $, which are ${\cal O}(\Lambda_S^4/M^2)$.
%%%%%%%%%%%%%%%%%%%%%%%%figure%%%%%%%%%%%%%%%%%%%%%%%%
\begin{figure}[t]  
%\psdraft
\vspace{1.cm} \centerline{
\psfig{figure=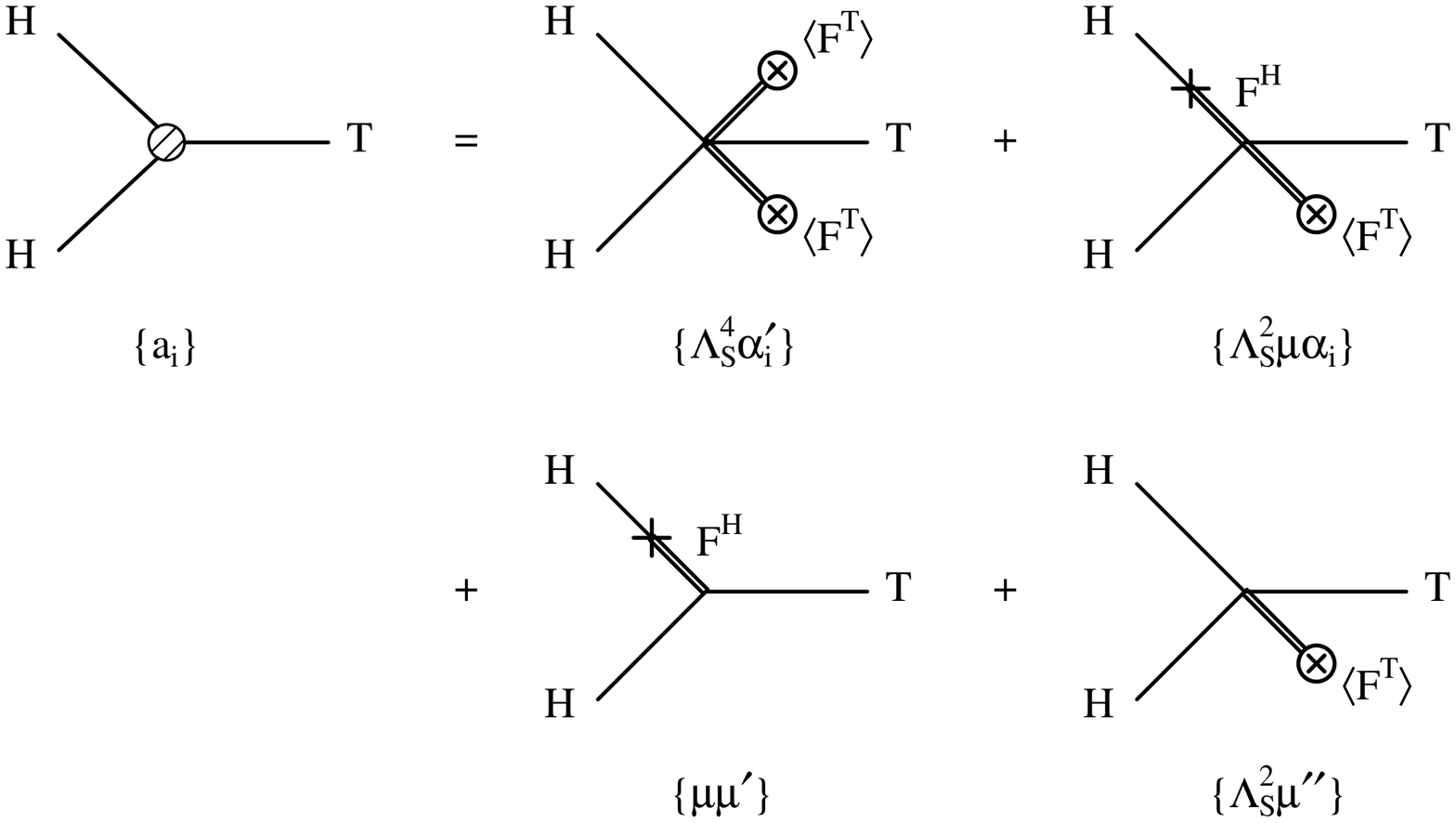,height=8cm,width=6cm,bbllx=10.cm,%
bblly=12.cm,bburx=19.cm,bbury=23.cm}}
\caption{\footnotesize Diagrammatic origin of different contributions
to the cubic $H^2T$ couplings [$a_i$ in Eq.~(\ref{ai})].}
\label{HHT}
\end{figure}
%%%%%%%%%%%%%%%%%%%%%%%%figure%%%%%%%%%%%%%%%%%%%%%%%%
The coefficients of the ${\cal O}(H^2)$ terms of $V$
are given by
\be m_1^2 =  |\mu|^2 -  \alpha_1 \tilde{m}^2 \; , \;\;\;\;
m_2^2 =  |\mu|^2 - \alpha_2 \tilde{m}^2 \; , \;\;\;\;
m_3^2 = \mu' \tilde{m}\ ,
\ee
and are generically ${\cal O}(\tilde{m}^2)={\cal O}(\Lambda_S^4/M^2)$.
The coefficients of the ${\cal O}(T H^2)$ terms of $V$
are given by
\bea a_1 = - {1\over M}(\alpha'_1\tilde{m}^2 +
\bar{\alpha}_3\mu\tilde{m}-\bar{\mu} \mu')  \ & , &  \;\;  
a_3 =  {\displaystyle{1\over M}} \tilde{m}\mu'' \ ,  
\nonumber\\
a_2 = - {\displaystyle{1\over M}}(\alpha'_2 \tilde{m}^2 +
\bar{\alpha}_3\mu\tilde{m} -\bar{\mu} \mu')  \ & , & \;\; 
a_4 = -{\displaystyle{1\over M}}[\alpha'_3\tilde{m}^2  
+(\alpha_1 + \alpha_2)\mu\tilde{m}]\ ,
\label{ai}
\eea 
and are generically  ${\cal O}(\tilde{m}^2/M)={\cal O}(\Lambda_S^4/M^3)$.
The origin of the different contributions to these cubic
couplings is traced back to the diagrams in Fig.~\ref{HHT}, which
carry self-explanatory labels. Double lines represent auxiliary fields
and  crossed circles represent the lowest order VEV 
$\langle F^T \rangle_0 = - \Lambda_S^2$.
%%%%%%%%%%%%%%%%%%%%%%%%figure%%%%%%%%%%%%%%%%%%%%%%%%
\begin{figure}[t]  
%\psdraft
\vspace{1.cm} \centerline{
\psfig{figure=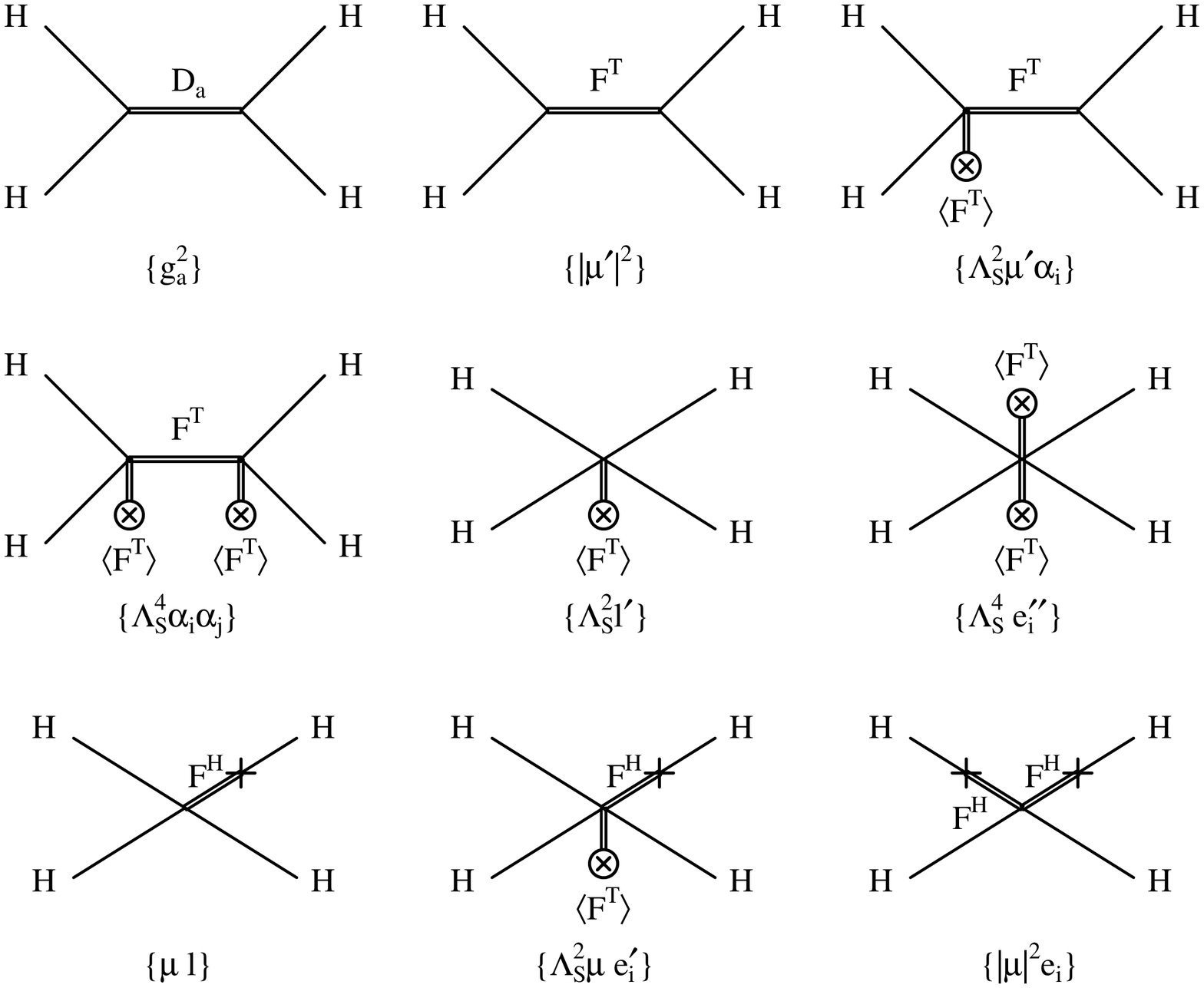,height=12cm,width=6cm,bbllx=10.cm,%
bblly=5.cm,bburx=20.cm,bbury=23.cm}}
\caption{\footnotesize Diagrammatic origin of different contributions
to the quartic Higgs couplings [$\lambda_i$ in Eqs.~(\ref{H4D}) and
(\ref{H4F})].}
\label{H4}
\end{figure}
%%%%%%%%%%%%%%%%%%%%%%%%figure%%%%%%%%%%%%%%%%%%%%%%%%

Finally, the coefficients of the ${\cal O}(H^4)$ terms of $V$ 
(quartic in the Higgs fields without $T$) receive two different 
types of contributions:  \be
\lambda_i = \lambda_i^{(D)} +  \lambda_i^{(F)}\ .  \ee The
$\lambda_i^{(D)}$ arise from $V_D$ as usual: \be
\lambda_1^{(D)}=\lambda_2^{(D)}={1\over 4}(g^2 + g_Y^2) \ , \;\;
\lambda_3^{(D)}={1\over 4}(g^2 - g_Y^2) \ , \;\;
\lambda_4^{(D)}=-{1\over 2}g^2 \ ,
\label{H4D}
\ee while $\lambda_5^{(D)}=\lambda_6^{(D)}=\lambda_7^{(D)}=0$.  The
$\lambda_i^{(F)}$ are the direct contributions from  the ${\cal
O}(H^4)$ terms in $V_F$:  \bea \lambda_1^{(F)} & = &  {1\over
M^2}\left[ \tilde{m}^2 (2 \alpha_1^2 - e''_1)  -2 \tilde{m} (\bar{\mu}
e'_6 + {\rm h.c.})  - 2 |\mu|^2 (e_3+e_4)\right]\ , \nonumber\\
\lambda_2^{(F)} & = &  {1\over M^2}\left[ \tilde{m}^2  (2 \alpha_2^2 -
e''_2)  -2 \tilde{m} (\bar{\mu} e'_7 + {\rm h.c.}) - 2 |\mu|^2
(e_3+e_4)\right]\ , \nonumber\\ \lambda_3^{(F)} & = & {1\over
M^2}\left[ \tilde{m}^2 (2 \alpha_1 \alpha_2 - e''_3)  -\tilde{m}
\left[\bar{\mu} (e'_6+e'_7) +{\rm h.c.}\right]  - |\mu|^2 (e_1+e_2+2
e_4)\right]\ , \nonumber\\ \lambda_4^{(F)} & = & -{1\over M^2}\left[
\tilde{m}^2 e''_4  + \tilde{m} \left[(e'_6+e'_7) \bar{\mu} +{\rm
h.c.}\right] + |\mu|^2 (e_1+e_2+ 2e_3) - |\mu'|^2\right]\ ,
\nonumber\\ \lambda_5^{(F)} & = & {\ell'\tilde{m}\over M} \ ,
\nonumber\\ \lambda_6^{(F)} & = & -{1\over M^2}\left[ \tilde{m}^2
e''_6   + \tilde{m} \left[\mu(e'_1+e'_3+e'_4)+ \mu' \alpha_1 \right]
\right] + { \ell\bar{\mu}\over M}\ , \nonumber\\ \lambda_7^{(F)} & =
& -{1\over M^2}\left[ \tilde{m}^2 e''_7   +\tilde{m}
\left[\mu(e'_2+e'_3+e'_4)+ \mu' \alpha_2 \right]\right] + {
\ell\bar{\mu}\over M}\ .
\label{H4F}
\eea Figure~\ref{H4} shows the diagrammatic origin of these terms,
again with labels as appropriate to identify different types of
contributions. The parameters $\lambda_i^{(F)}$ are generically 
${\cal O}(\tilde{m}^2/M^2)={\cal O}(\Lambda_S^4/M^4)$. Note that these 
`hard-breaking' terms do not spoil the stability of the electroweak scale
since the cut-off of the effective theory is of ${\cal O}(M)$.
For completeness, we recall that quartic couplings also receive
sizeable radiative corrections even in the conventional MSSM scenario,
as is well known \cite{loop}.

As already mentioned, the terms of $V(T,H_1,H_2)$ shown 
above are sufficient to compute VEVs and masses at
lowest orders in the $v/M$ expansion.
The general qualitative features of the results can
be easily inferred. The minimization conditions of $V$ 
give constraints on the Higgs VEVs\footnote{Equivalently,
such constraints can be interpreted as tuning conditions
on the mass parameters $m_i^2$, {\it i.e.} on the parameters
they contain. As mentioned in subsection 3.1, the presence 
of sizeable ${\cal O}(\Lambda_S^4/M^4)$ quartic couplings can
alleviate the required amount of fine tuning, 
for fixed ${\tilde m}> v$.}
and produce a small VEV for $T$, {\it i.e.} $\langle T \rangle=
{\cal O}(v^2/M)$, induced by the ${\cal O}(T^2)+{\cal O}(T H^2)$ 
terms of $V$. 
The \SUSY scale is $F^2 = \langle V \rangle 
= \Lambda_S^4 + {\cal O}(\lambda v^4)$,
where $\lambda$ stands for either $g^2$ or 
$\Lambda_S^4/M^4={\tilde m}^2/M^2$.
In the limit of $CP$ conservation, 
the physical spectrum contains a pair of charged 
Higgs bosons, three $CP$-even neutral bosons
and two $CP$-odd neutral bosons.
The $T$-$H$ mixing angles of the neutral sectors
are generically ${\cal O}(v/M)$, so in each sector 
one mass eigenstate is mainly $T$-like (singlet)
whereas the other(s) are mainly Higgs-like.
The mass eigenvalues can be computed by perturbative
diagonalization of the mass matrices and have the 
parametric form ${\cal O}(\tilde{m}^2) + {\cal O}(\lambda v^2)+ \ldots$.
The leading term ${\cal O}(\tilde{m}^2)$ is absent in at 
least one Higgs-like eigenvalue of the neutral sector (this is generic, 
see {\it e.g.} \cite{CE}), 
and may be absent in other eigenvalues in specific models. 
The ${\cal O}(\lambda v^2)$ terms arise from several 
sources, including kinetic normalization\footnote{
Kinetic normalization is easily deduced from 
eq.~(\ref{lkine}).
Incidentally, notice that $T$-$H$ kinetic mixing arises 
at ${\cal O}(v/M)$ in general coordinates, hence it can
contribute to ${\cal O}(\lambda v^2)$ mass terms.
In normal coordinates, however, such a mixing only arises 
at ${\cal O}(v^3/M^3)$, so it can be neglected. We add here another 
minor comment, concerning the ${\cal O}(\lambda v^2)$ corrections
to the $T$ mass eigenvalues $m^2_{T_{1,2}}$. 
Part of such corrections originate from ${\cal O}(T^3)$ 
and ${\cal O}(T^2 H^2)$ terms of $V$. The full computation 
of these terms, which we have not presented, is straightforward. 
To this purpose, for completeness one should also take into 
account a few higher order terms of $K$ and $W$ not explicitly shown 
in eqs.~(\ref{wnor},\ref{knor}).}.

It is worthwhile mentioning that the effects of the $T$ field
on the Higgs VEVs and masses could also be studied by
a slightly different (albeit essentially equivalent) 
approach, {\it i.e.} one can integrate out the $T$ scalars 
from the beginning. This operation effectively reduces 
$V(T,H_1,H_2)$ to a simpler potential $V(H_1,H_2)$,
a function of the Higgs doublets only.
The potential $V(H_1,H_2)$ 
contains additional effective quartic terms, obtained by contractions
of the original ${\cal O}(T H^2)$ cubic terms through the exchange
of the massive $T$ field. Thus the coefficients $\lambda_i$ 
of ${\cal O}(H^4)$ terms receive additional 
effective contributions $ \delta \lambda_i^{(F)}$,
whose expression is:
\bea 
\delta \lambda_1^{(F)} & = &
\frac{1}{(\alpha_t^2 - |\rho_t|^2){\tilde m}^2 }
\left[  - 2\alpha_t |a_1|^2  + \left(
{\bar\rho}_t a_1^2 +{\rm h.c.} \right) \right] \ ,
\nonumber\\ 
\delta \lambda_2^{(F)} & = &  
\frac{1}{(\alpha_t^2 - |\rho_t|^2){\tilde m}^2 }
\left[  - 2\alpha_t |a_2|^2  + \left(
{\bar\rho}_t a_2^2 +{\rm h.c.} \right) \right] \ ,
\nonumber\\ 
\delta \lambda_3^{(F)} & = &  
\frac{1}{(\alpha_t^2 - |\rho_t|^2){\tilde m}^2 }
\left[  - \alpha_t (a_1 \bar{a}_2 + a_2 \bar{a}_1) 
+ \left({\bar\rho}_t  a_1 a_2 +{\rm h.c.} \right) \right] \ , 
\nonumber\\ 
\delta \lambda_4^{(F)} & = &  
\frac{1}{(\alpha_t^2 - |\rho_t|^2){\tilde m}^2 }
\left[ - \alpha_t (|a_3|^2 + |a_4|^2 ) 
+ \left({\bar\rho}_t a_3 \bar{a}_4 +{\rm h.c.} \right) \right] \ , 
\nonumber\\ 
\delta \lambda_5^{(F)} & = &  
\frac{1}{(\alpha_t^2 - |\rho_t|^2){\tilde m}^2 }
\left[  - 2 \alpha_t a_3 a_4  
+{\bar\rho}_t a_3^2 + \rho_t a_4^2  \right] \ , 
\nonumber\\ 
\delta \lambda_6^{(F)} & = &
\frac{1}{(\alpha_t^2 - |\rho_t|^2){\tilde m}^2 }
\left[  - \alpha_t ( \bar{a}_1 a_3 + a_1 a_4)
+{\bar\rho}_t a_1 a_3 + \rho_t \bar{a}_1 a_4 \right] \ , 
\nonumber\\ 
\delta \lambda_7^{(F)} & = &
\frac{1}{(\alpha_t^2 - |\rho_t|^2){\tilde m}^2 }
\left[  - \alpha_t ( \bar{a}_2 a_3 + a_2 a_4)
+{\bar\rho}_t a_2 a_3 + \rho_t \bar{a}_2 a_4 \right] \ .
\label{Tex}
\eea
%%%%%%%%%%%%%%%%%%%%%%%%figure%%%%%%%%%%%%%%%%%%%%%%%%
\begin{figure}[t]  
%\psdraft
\vspace{1.cm} \centerline{
\psfig{figure=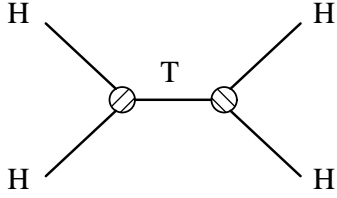,height=2cm,width=6cm,bbllx=3.cm,%
bblly=20.cm,bburx=12.cm,bbury=23.cm}}
\caption{\footnotesize Diagrammatic origin of the $T$-exchange
contributions to the Higgs quartic coupling [$\delta\lambda_i^{(F)}$
in Eq.~(\ref{Tex})]. The blob is defined by figure~\ref{HHT}.}
\label{Texch}
\end{figure}
%%%%%%%%%%%%%%%%%%%%%%%%figure%%%%%%%%%%%%%%%%%%%%%%%% 
The diagrammatic origin of these terms is depicted in Fig.~4, with
the blobs representing  the ${\cal O}(TH^2)$ couplings, as schematically 
given by  Fig.~2. Notice that the parameters $\delta \lambda_i^{(F)}$ 
are formally of the same order as the parameters  $\lambda_i^{(F)}$,
{\it i.e.} ${\cal O}(\tilde{m}^2/M^2)={\cal O}(\Lambda_S^4/M^4)$.
The minimization of the reduced potential $V(H_1,H_2)$ gives 
the same conditions on the Higgs VEVs that are obtained by
minimizing the full $V(T,H_1,H_2)$, as it should, and
the Higgs boson mass eigenvalues are approximately 
reproduced\footnote{More precisely, the same results of
the full approach are obtained for Higgs bosons whose mass 
is lighter than the $T$ mass, {\it e.g.} ${\cal O}(\lambda v^2)$ rather
than ${\cal O}(\tilde{m}^2)$, or Higgs bosons of any mass
that are not mass-mixed with $T$, {\it e.g.} the charged one
and possibly some neutral one. On the other hand, 
if a neutral Higgs boson has both an ${\cal O}(\tilde{m}^2)$ 
leading mass comparable to that of $T$ and mass-mixing with 
$T$, then the ${\cal O}(\lambda v^2)$ corrections to 
its mass induced by $T$-$H$ mixing are only approximately 
reproduced by this method.  In the latter case, if one
is interested in those ${\cal O}(\lambda v^2)$ corrections, 
the full $V(T,H_1,H_2)$ should be used.}.

To conclude this discussion, we note again that the limit 
$\tilde{m}/M=\Lambda_S^2/M^2 \rightarrow 0$,  keeping 
$\tilde{m}=\Lambda_S^2/M$  
fixed, corresponds to a standard MSSM scenario with
conventional soft terms, and the field $T$ decoupled from the
observable matter.  Here, however, we are interested in the opposite 
limit, in which $\tilde{m}/M=\Lambda_S^2/M^2$ is not negligible. 
In this case, as anticipated, the Higgs couplings deviate from the
usual MSSM values, which can have a significant impact for the SUSY Higgs 
sector phenomenology. It is important to stress also that, 
although similar deviations have been reported previously
in the literature (see {\it e.g.} \cite{Polonsky}),
our analysis includes all the relevant effects, some of which have 
not been considered by other studies. Indeed, it is common 
to treat the superfield $T$ simply as the source of SUSY breaking, 
providing a non-zero $\langle F^T\rangle$, which then generates 
different effects, whereas other contributions of comparable 
importance, which come from the degrees of freedom associated to
$T$, are often neglected. Examples of the latter effects,
included in our analysis above, are the contributions to the 
Higgs potential that come from $F^T$ exchange, or from $T$ 
exchange ({\it i.e.}, equivalently, from $T$-$H$ mixing effects).
Finally we recall again that, to compute the spectrum and
the self-interactions of Higgs and $T$ fields, both
the scalar potential and the derivative terms of 
eq.~(\ref{lkine}) should be taken into account.

\subsection{The neutralino/goldstino sector and the chargino sector}

Another sector of the theory that changes with respect to the 
conventional MSSM is the neutralino sector. In particular,
the fermionic partner of $T$ can in principle mix with the 
Higgsinos and gauginos after electroweak symmetry breaking. 
We will present here the neutralino and
chargino mass matrices at ${\cal O}(v^2)$, specializing 
the general expression (\ref{generalM}) of the fermion mass 
matrix and taking into account kinetic normalization.
Since we have already shown the explicit expansions of $K$ and $W$ 
in normal coordinates, eqs.~(\ref{wnor},\ref{knor}), 
we only need to add the analogous expansion of $f_{ab}$, up to 
${\cal O}(H^2)$. To this purpose, it is sufficient
to expand in $T$ the expressions of $f_{ab}^{(s)}$ (singlet) 
and $f_{ab}^{(t)}$ (triplet)
already given in eqs.~(\ref{fsin},\ref{ftri}):
\bea 
f_{ab}^{(s)} &=&
{\delta_{ab} \over g_a^2} \left[ 1 + 2 \eta_a {T \over M} 
+ 2 \eta_a' {T^2 \over M^2} + \ldots  
+\! \left( h_a+h_a' {T \over M} + \ldots  \right) \!
{H_1\cdot H_2\over M^2}+ \ldots \right]
\\ 
f_{AY}^{(t)} &=&
{1\over g g_Y }\left(\omega + \omega' {T\over  M}  + \ldots   \right)
{H_1 \cdot \sigma^A H_2 \over M^2} 
+ \ldots
\eea
where inverse powers of (zero-th order) gauge
couplings have been inserted for convenience.
As already mentioned, the fermionic kinetic terms are no 
longer canonical after electroweak breaking.
The fermionic mass matrices we present below are
already referred to the canonical basis, {\it i.e.} the
symbols $\lambda,\tilde{H},\tilde{T}$ will denote fields 
that are already canonically normalized. 
Moreover, for simplicity we will assume that
all parameters in $W$, $K$ and $f_{ab}$ are real.

The $5\times 5$ neutralino mass matrix, in the basis 
$(\lambda_B^0, \lambda_W^0,\tilde{H}_1^0, \tilde{H}_2^0,
\tilde{T})$, reads
\bea
\label{MNnorcoord}
&& {\cal M}_N =  \left(
\begin{array}{ccccc}
M_1 & \displaystyle{\kappa_{\omega} v^2 \over 2 M^2} & 
-M_Z s_w c_\beta &  M_Z s_w s_\beta
&\displaystyle{g_Y\eta_B\over 4\sqrt{2}M}v^2 c_{2\beta}   \\
& & & & \\ 
\displaystyle{\kappa_{\omega} v^2 \over 2 M^2} & M_2 &  
M_Z c_w c_\beta  &-M_Z c_w s_\beta  &
\displaystyle{-g\eta_W\over 4\sqrt{2}M}v^2 c_{2\beta} \\ 
& & & & \\ 
-M_Z s_w c_\beta &  M_Z c_w c_\beta & 
\displaystyle{\kappa_1 v^2 \over 2 M^2} & \mu_N &  
\displaystyle{- \mu^2 \over \sqrt{2}\Lambda_S^2} v c_\beta \\ 
& & & & \\ 
M_Z s_w s_\beta & -M_Z c_w s_\beta & \mu_N & 
\displaystyle{\kappa_2 v^2 \over 2 M^2} &  
\displaystyle{- \mu^2 \over \sqrt{2}\Lambda_S^2} v s_\beta \\
& & & & \\ 
\displaystyle{g_Y\eta_B\over 4\sqrt{2}M}v^2 c_{2\beta} &
\displaystyle{-g\eta_W\over 4\sqrt{2}M}v^2 c_{2\beta} &  
\displaystyle{- \mu^2 \over \sqrt{2}\Lambda_S^2} v c_\beta &
\displaystyle{- \mu^2 \over \sqrt{2}\Lambda_S^2} v s_\beta & 
\displaystyle{\mu^3 \over 2\Lambda_S^4} v^2 s_{2\beta} 
\end{array}
\right), \nonumber\\ && \eea
where
\bea 
\label{Mi} 
M_a & = & \eta_a\tilde{m} 
-{v^2 \over 2 M^2} \left[
{\tilde m} \left( \eta_a (\alpha_1 c_\beta^2+\alpha_2 s_\beta^2)
+ {1 \over 4} (2 h_a \eta_a - h'_a) s_{2\beta}
+ 2 ( \eta_a^2 - \eta'_a) \xi_t \right) \right.
\nonumber\\  
& - & \left.
{1 \over 2} (\mu h_a + \mu' \eta_a s_{2\beta}) \right] \ ,
\\
\label{muN}
\mu_N & = & \mu
- {v^2\over 2 M^2} \left[ \mu 
\left( e_1 c_\beta^2 +e_2 s_\beta^2+{3\over 2}(e_3+e_4)\right)
- \mu' \xi_t - \ell M s_{2\beta}
\right. 
\nonumber\\  
&+& \left.
{\tilde m} \left(
{1 \over 2}(e'_3+ e'_4) s_{2\beta} +
2 (e'_6 c_\beta^2 + e'_7 s_\beta^2) + \alpha_3 \xi_t \right)
\right] \ ,
\\
\kappa_{\omega} & = &
- {1 \over 2}\mu\omega + {1\over 4} {\tilde m} 
[ \omega(\eta_1+\eta_2) - \omega' ] s_{2\beta} \ ,
\\
\kappa_1 &=&
- {1 \over 2}\mu( 2 e_1 + e_3+e_4)s_{2\beta} + \ell M s_\beta^2
- {\tilde m}(e'_1 c_\beta^2 + e'_6 s_{2\beta}) \ ,
\\ 
\kappa_2 &=&
- {1 \over 2}\mu( 2 e_2 + e_3+e_4)s_{2\beta} + \ell M c_\beta^2
- {\tilde m}(e'_2 s_\beta^2 + e'_7 s_{2\beta}) \ .
\eea
The auxiliary parameter $\xi_t$ is defined by 
$\langle T \rangle = \xi_t v^2/2M$ and is related to the 
other parameters by the minimization condition 
$\langle \partial_T V \rangle=0$ of the potential 
of eq.~(\ref{Vnorcoord}). At lowest order:
\be
\xi_t = {1 \over \alpha_t + \rho_t}
\left\{ \left( \alpha'_1 c_\beta^2 + \alpha'_2 s_\beta^2
+ {1 \over 2} \alpha'_3 s_{2\beta} \right)
+ {\mu \over {\tilde m}} 
\left[ {1 \over 2}(\alpha_1 + \alpha_2) s_{2\beta} 
+ \alpha_3 \right]
- {\mu'' \over 2 {\tilde m}} s_{2\beta} 
- {\mu \mu' \over {\tilde m}^2 }
\right\}\ .
\ee
We have also exploited the conditions $\langle \partial_T V \rangle=0$ 
and $\langle \partial_{H_i^0} V \rangle=0$ at lowest order\footnote{
The conditions $\langle \partial_{H_i^0} V \rangle=0$ imply
$(\alpha_1 {\tilde m}^2 - \mu^2) c_\beta - \mu' {\tilde m} s_\beta
= {\cal O}(v^2)$ and
$(\alpha_2 {\tilde m}^2 - \mu^2) s_\beta - \mu' {\tilde m} c_\beta
= {\cal O}(v^2)$.} 
to simplify the $\tilde{T}$-$\tilde{T}$ and $\tilde{H}^0$-$\tilde{T}$ 
entries of the mass matrix ${\cal M}_N$ above.

Among the five neutralinos, four are massive and 
one corresponds to the massless goldstino. 
The leading terms in ${\cal M}_N$ are ${\cal O}({\tilde m})$
and appear in the entries $\lambda_B^0$-$\lambda_B^0$, 
$\lambda_W^0$-$\lambda_W^0$ and $\tilde{H}_1^0$-$\tilde{H}_2^0$,
as usual. Therefore the four massive eigenstates have  
dominant gaugino or Higgsino components, whereas
the goldstino has a dominant ${\tilde T}$ component.
At linear order in $v$, we find terms of two types:
the usual ${\tilde H}^0$-$\lambda^0$ mixing terms, which 
are ${\cal O}(M_Z)$, and ${\tilde H}^0$-${\tilde T}^0$ 
mixing terms, which are ${\cal O}(v{\tilde m}/M)={\cal O}
(v \Lambda_S^2/M^2)$. Notice that 
the latter can be larger than the former if ${\tilde m}/M
=\Lambda_S^2/M^2$ is sizeable\footnote{This is somewhat 
reminiscent of the situation in the scalar potential, 
where the couplings $\lambda_i^{(F)}$ could be more important 
than the usual $\lambda_i^{(D)}$ couplings.}, {\it i.e.}
Higgsinos could have larger mixing with ${\tilde T}$ than
with gauginos. At ${\cal O}(v^2)$, other effects appear.

Let us now consider the approximate identification of 
the goldstino. If ${\cal M}_N$ were computed exactly, 
it would have an exactly massless eigenvalue.
In fact, it is straightforward to verify that the mass matrix 
(\ref{MNnorcoord}) approximately annihilates the vector 
\be
\label{vect}
\left({g_Y \over 4\sqrt{2}} v^2 c_{2\beta},
-{g \over 4\sqrt{2}} v^2 c_{2\beta},
-{1 \over \sqrt{2}} \mu v s_\beta,
-{1 \over \sqrt{2}} \mu v c_\beta, - \Lambda_S^2 \right)^T \, .
\ee
This is consistent with the general properties recalled in sect.~2.1, 
since the entries of the vector (\ref{vect}) contain the lowest 
order VEVs of the (canonically normalized) auxiliary fields. 
Thus the explicit form of the goldstino field $\tilde{G}$ is
\be
\label{gcomp}
\tilde G \simeq
\left( 1 - { \mu^2 v^2 \over 4\Lambda_S^4} \right)
\tilde{T}
+ {\mu v \over \sqrt{2} \Lambda_S^2 }  
(s_\beta \tilde{H}_1^0 +c_\beta \tilde{H}_2^0) 
+ {v^2  c_{2 \beta} \over 4 \sqrt{2} \Lambda_S^2 }  
(- g_Y \lambda_B^0 + g \lambda_W^0)  
\ ,  
\ee
up to an overall phase and up to higher order terms in $v$.
Notice that the gaugino combination in eq.~(\ref{gcomp})
is a $Z$-ino, and that its coefficient vanishes if
the $D$-terms have vanishing VEVs, {\it i.e.} for 
$|\tan\beta|=1$.

The chargino sector contains the same degrees of freedom as in the
MSSM ($\lambda^{\pm}, \tilde{H}^\pm$) and the chargino mass matrix 
has the same form: 
\bea
\label{MCHnorcoord}
&& {\cal M}_C =  \left(
\begin{array}{cc}
M_{2} & \sqrt{2}M_W s_\beta \\  \sqrt{2}M_W c_\beta & -\mu_C
\end{array}
\right) \, . 
\eea 
The parameter $M_2$ is the same one that appears in the neutralino 
mass matrix and is given by eq.~(\ref{Mi}), whereas 
the parameter $\mu_C$ is different from $\mu_N$ of
eq.~(\ref{muN}), and is given by
\bea
\mu_C & = & \mu
- {v^2\over 2 M^2} \left[ { 1 \over 2} \mu 
( e_1 c_\beta^2 +e_2 s_\beta^2+ e_3 + 2 e_4)
-  \mu' \xi_t - {1\over 2} \ell M s_{2\beta}
\right. 
\nonumber\\  
&+& \left.
{\tilde m} \left(
{1 \over 2} e'_4 s_{2\beta} +
e'_6 c_\beta^2 + e'_7 s_\beta^2 + \alpha_3 \xi_t \right)
\right] \ ,
\eea

One of the main effects of electroweak breaking is to 
lift the zero-th order mass-degeneracy of the 
three Higgsino-like states (two neutral and one 
charged). Part of this lifting originates from the 
usual ${\tilde H}$-$\lambda$ entries, which induce
${\cal O}(M_Z^2/{\tilde m}^2)$ relative splittings.
On top of that, we see that the effective non-renormalizable 
operators generate ${\cal O}(v^2/M^2)$ 
relative splittings, which could be comparable to the standard 
ones. In particular, $\mu_C$ is split from $\mu_N$,
and splitting effects arise also within the neutral 
sector, either from $\tilde{H}^0_i$-$\tilde{H}^0_i$
entries (proportional to $\kappa_i$) or 
from $\tilde{H}^0_i$-$\tilde{T}$ entries. 
These effects, which regard the Higgsino sector
and are not related to Higgsino-gaugino mixing, can be compared 
with analogous ones that are generated at one-loop level
in the MSSM (see {\it e.g.} \cite{Giudice:1995qk}).
In the latter case, the induced relative splittings 
scale as $v^2/ {\tilde m}^2 $, times
a $1/16 \pi^2$ loop factor.

\section{Simple examples}
\setcounter{equation}{0}
\renewcommand{\theequation}{5.\arabic{equation}}

For illustrative purposes we devote this section to 
present two simple examples, {\it i.e.} two models with 
a small number of parameters. For simplicity we choose both models to be 
symmetric under exchange of $H_1$ and $H_2$, in spite of which,
vacua with $\tan\beta\neq 1$ can still be achieved, as we will explicitly 
show (of course, in models which are not symmetric it is trivial to 
obtain $\tan\beta\neq 1$). 
Some general results concerning electroweak breaking in the case 
of symmetric potentials are collected in Appendix A.

\subsection*{Example A}

Our first example is a model which can accomodate both $\tan\beta=1$ 
and $\tan\beta \neq 1$ (depending on the choice of parameters), 
even though there is a symmetry $H_1\leftrightarrow H_2$.
The model is written in normal coordinates, so we can
specialize the general results obtained in sect.~4.
The superpotential, gauge kinetic functions and K\"ahler potential are
chosen as 
\be W =
\Lambda_S^2 T + \mu H_1\cdot H_2 + {\ell\over 2M}(H_1\cdot H_2)^2 \ ,
\;\;\;\;  f_{ab} = {\delta_{ab}\over g_a^2} \left(1 + 2 {\eta_a \over
M} T \right)\ , \ee and \bea K & = & |T|^2 +  |H_1|^2 +  |H_2|^2
\nonumber\\ & - & {\alpha_t \over 4 M^2} |T|^4 + {\alpha_1 \over
M^2}|T|^2  \left(|H_1|^2+|H_2|^2\right) + {e_1 \over
2M^2}\left(|H_1|^4+|H_2|^4\right)\ ,  \eea
where all parameters are taken to be real, with $\alpha_t>0$.
We will sometimes use the auxiliary parameter  
${\tilde m}=\Lambda_S^2/M$. 

We will analyse the model perturbatively in the Higgs VEVs, 
following the general discussion of the previous sections.
We will only retain the first terms of the expansion,
which will be sufficient to illustrate the main qualitative
features of this example.
The results can easily be obtained by specializing the general 
formulae presented above or, equivalently, by direct computation.

At zero-th order, {\it i.e.} for vanishing Higgs VEVs, we have
$\langle T\rangle_0=0$, SUSY is broken by  $\langle F^T \rangle_0 
= -\Lambda_S^2$, $\tilde{T}$ is the goldstino and
the complex $T$ field has mass $m_T^2 = \alpha_t\tilde{m}^2$. 
The effects of electroweak breaking start to appear at next order,
{\it i.e.} when the potential $V(T,H_1,H_2)$ is minimized and
the Higgses take VEVs. In particular, since 
$V(T,H_1,H_2)$ contains ${\cal O}(T H^2)$ cubic 
terms\footnote{Notice that the only non-vanishing 
coefficient of ${\cal O}(T H^2)$ 
terms in (\ref{ai}) is $a_4=-2\alpha_1\mu\tilde{m}/M$.}, 
$T$ receives a small induced VEV $\langle T\rangle=\alpha_1\mu 
v_1v_2/(\alpha_t\Lambda_S^2)$ and $T$-$H$ mass mixing appears.
Instead of keeping the field $T$ together with the Higgses, 
however, we find it more convenient to use the alternative
method mentioned in the previous section, {\it i.e.} to
integrate out $T$ and study a reduced effective
potential for the Higgs doublets only. This choice
is also supported by the special fact that all Higgs boson 
masses turn out to be ${\cal O}(\lambda v^2)$ in this
model, {\it i.e.} naturally lighter than the $T$ mass, which is 
${\cal O}({\tilde m}^2)$. 

The Higgs VEVs and spectrum are determined by an effective 
quartic potential $V(H_1,H_2)$ with particular values
for its mass terms: \be m_1^2=m_2^2=\mu^2-\alpha_1 \tilde{m}^2\
,\;\;\;\; m_3^2=0\ , \ee and quartic couplings\footnote{
The only contribution induced by $T$-exchange is the
term proportional to $1/\alpha_t$ in $\lambda_4$,
as can be checked from the general formulae
(\ref{Tex}).}
\bea
\lambda_1=\lambda_2&=&{1\over 4}(g^2+g_Y^2)+2\alpha_1^2{\tilde{m}^2\over
M^2}\ ,\nonumber\\ \lambda_3&=&{1\over 4}(g^2-g_Y^2)+{2\over
M^2}(\alpha_1^2\tilde{m}^2-e_1\mu^2)\ ,\nonumber\\  \lambda_4&=&-{1\over
2}g^2-2\left(e_1+2{\alpha_1^2\over \alpha_t}\right){\mu^2\over M^2}\
,\nonumber\\  \lambda_5&=&0\ ,\nonumber\\
\lambda_6=\lambda_7&=&{\ell\mu\over M}\ .  
\eea
This example is represented by point A in Fig.~\ref{thdmbreak},  
although now we are in  the extreme case $m_1^2/|m_3^2|\rightarrow 
-\infty$ and this ratio is finite  in the figure. A correct electroweak
breaking can nevertheless be achieved.  
We can apply the general formulae given in Appendix A to write down the
minimization conditions that give $v^2$ and $\sin 2\beta$,
as well as the expressions of the Higgs masses.  
Concerning the value of $\tan\beta$, we have the two possible
solutions   \be |\tan\beta| = 1 \ , \ee and 
\be \sin2\beta={\ell\mu/M\over
({g^2+g_Y^2})/4+2\hat{e}_1\mu^2/M^2}\ ,
\label{tbneq1}
\ee where we use $\hat{e}_1\equiv
e_1+\alpha_1^2/\alpha_t$.  Both solutions are possible depending on
the choice of parameters, as explained in Appendix A, and
in both cases ${\rm sgn}(\tan\beta)= -{\rm sgn}(\ell\mu/M)$.
It is not restrictive to take $\ell\mu/M <0$, so that
$\tan\beta >0$. Using this convention, the explicit expressions 
for the Higgs masses are the following.  
\be
\begin{array}{ll}
\hspace{2.5cm}{\underline{\tan\beta=1}}:&
\hspace{2.5cm}{\underline{\tan\beta\neq 1}}:\\ &\\
m_h^2={\displaystyle 2\left(\alpha_1^2{\tilde{m}^2\over
M^2}-\hat{e}_1{\mu^2\over M^2} +{\ell\mu\over M}\right)v^2}\ , &
m_h^2={\displaystyle \left[{1\over
4}(g^2+g_Y^2)+2\alpha_1^2{\tilde{m}^2\over M^2}+{\ell\mu\over
M}s_{2\beta}\right]v^2}\ ,\vspace{0.2cm}\\
m_H^2={\displaystyle\left[{1\over 4}(g^2+g_Y^2)+2\hat{e}_1{\mu^2\over
M^2} -{\ell\mu\over M}\right]v^2}\ ,&m_H^2={\displaystyle  -\left[{1\over
4}(g^2+g_Y^2)+2\hat{e}_1{\mu^2\over M^2}\right]v^2c^2_{2\beta}}\
,\vspace{0.2cm}\\ m_A^2={\displaystyle -{\ell\mu\over M} v^2 }\ ,&
m_A^2={\displaystyle  -\left[{1\over
4}(g^2+g_Y^2)+2\hat{e}_1{\mu^2\over M^2}\right]v^2}\
,\vspace{0.2cm}\\	 m_{H^\pm}^2={\displaystyle \left[{1\over
4}g^2+(2\hat{e}_1-e_1){\mu^2\over M^2} -{\ell\mu\over M}\right]v^2}\ , &
m_{H^\pm}^2={\displaystyle  -\left({1\over 4}g_Y^2+e_1{\mu^2\over
M^2}\right)v^2}\ .
\end{array}
\ee 
We have used the general formulae of Appendix A, plus
eq.~(\ref{tbneq1}) to simplify $m_h^2$ in the case with
$\tan\beta\neq 1$. Notice that acceptable solutions with 
$\tan\beta=1$
can be obtained even if we set $e_1=0$, which further 
simplifies the model. To obtain solutions with $\tan\beta \neq 1 $,
however, we need $e_1 < 0$. Also notice that, in the phase
with  $\tan\beta \neq 1 $, the value of $\tan\beta$ is only 
determined up to an inversion ($\tan\beta \leftrightarrow 1/\tan\beta$),
which in fact leaves the spectrum invariant. This is a consequence
of the original discrete symmetry, and we can conventionally 
take $\tan\beta \geq 1$. 

%%%%%%%%%%%%%%%%%%%%%%%%figure%%%%%%%%%%%%%%%%%%%%%%%%
\begin{figure}[t]
%\psdraft
\centerline{
\psfig{figure=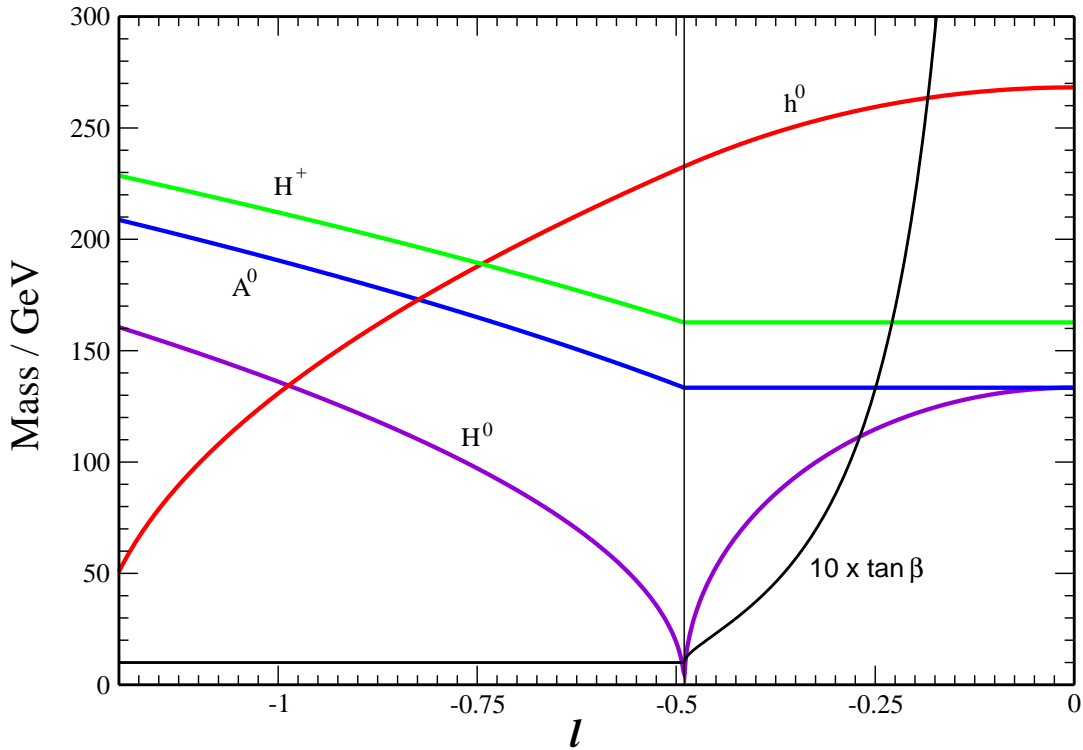,height=10cm,width=10cm,angle=-90,bbllx=3.cm,%
bblly=5.cm,bburx=20.cm,bbury=22.cm}}
\caption
{\footnotesize Higgs spectrum of example~A as a function of the
superpotential parameter $\ell$. Also shown is $\tan\beta$ (scaled by a
factor 10).}
\label{especA}
\end{figure}
%%%%%%%%%%%%%%%%%%%%%%%%%figure%%%%%%%%%%%%%%%%%%%%%%%%

In Fig.~5 we show a numerical example where both phases of the model 
are visible. We have fixed $\mu/M=0.6$,
$e_1=-1.3$, $\tilde{m}/M=0.5$, $\alpha_t=3.0$, $\alpha_1\simeq
\mu^2/\tilde{m}^2+\epsilon^2$ (with $0<\epsilon^2\ll 1$) and vary
$\ell$. For each parameter choice, the overall mass scale $M$ is adjusted
so as to get the right value of $v=246$ GeV. The closer $\alpha_1$ is
to $\mu^2/\tilde{m}^2$ the larger  $M/v$ can be. The figure shows the
Higgs spectrum and the parameter $\tan\beta$ (scaled by a factor 10
for clarity) as a function of the coupling $\ell$.
For $\ell\leq \ell_0$, with
\be \ell_0\equiv {M\over \mu}\left[{1\over
4}(g^2+g_Y^2)+2\hat{e}_1{\mu^2\over M^2} \right]\ , \ee  the minimum
lies at $\tan\beta=1$, while for $\ell > \ell_0 $, $\tan\beta$ 
increases with
$\ell$. For the choice of parameters used in this figure, $\ell_0\simeq
-0.49$. The spectrum is continuous across the critical value $\ell_0$,
although the mass of the `transverse' Higgs, $H^0$, goes through zero,
as was to be expected on general grounds for symmetric potentials.
We see that, except in the
neighbourhood of $\ell_0\simeq -0.49$ or for too negative values of 
$\ell$, the Higgs masses are sufficiently large to escape all current
experimental bounds (which are also lower than usual due to singlet
admixture, although this is typically a small effect). There is even a 
region of parameters for which $h^0$  is
the heavier of the Higgses, beyond the usual limit of $m_{h^0}\simlt
200$ GeV \cite{EQ} that applies in generic SUSY models only when they
are perturbative up to the GUT scale.

One may also be interested in finding  ${\cal O}(v^2)$
corrections to the $T$ mass, which is $\alpha_t {\tilde m}^2$
at leading order. To do this, we should reconsider 
the potential prior to the elimination of $T$, 
including ${\cal O}(T H^2)$ and ${\cal O}(T^2 H^2)$ 
terms. Taking also into account kinetic normalization, 
we get 
\be
m_T^2 = \alpha_t {\tilde m}^2 - {\alpha_1 v^2 \over 2 M^2} 
\left[  {\tilde m}^2 (3 \alpha_t - \alpha_1)
+ \mu^2 \left( 1 - 4 {\alpha_1 \over \alpha_t} \right) \right] \, .
\ee

In the neutralino sector, the goldstino $\tilde{G}$ 
is mainly ${\tilde T}$. As a result of electroweak breaking, 
$\tilde{G}$ also has small components along Higgsinos and 
(for $\tan\beta \neq 1$) along gauginos, see eq.~(\ref{gcomp}). 
The full $5\times 5$ neutralino mass matrix 
[including up to ${\cal O}(v^2/M)$ terms] is of the form 
(\ref{MNnorcoord}) with $\kappa_\omega=0$, 
$\kappa_1= \ell M s_\beta^2 - e_1 \mu s_{2\beta} $, 
$\kappa_2= \ell M c_\beta^2 - e_1 \mu s_{2\beta} $, 
\be
M_a=\eta_a\tilde{m} \left( 1-{\alpha_1 v^2\over 2 M^2}\right)
-{\eta_a^2 \alpha_1\mu v^2 \over  \alpha_t
M^2} s_{2\beta} \ ,
\ee
and
\be
\mu_N=\mu\left(1-{e_1 v^2\over 2M^2}\right)
+{\ell v^2\over 2M} s_{2\beta} \ .
\ee
The chargino sector is like in the MSSM, with
a mass matrix of the form (\ref{MCHnorcoord}) and 
\be
\mu_C=\mu\left(1-{e_1 v^2\over 4M^2}\right)
+{\ell v^2\over 4M} s_{2\beta} \ .  \ee

\subsection*{Example B}

Our general discussion and our previous example indicate 
that $T$-$H$ mixing effects generically 
arise after electroweak breaking, both in the scalar
and in the fermion sector. This does not exclude the 
possibility to construct models
where such mixing effects are absent and the 
goldstino remains a pure singlet ({\it i.e.} $\tilde{T}$), 
despite electroweak breaking. 
Here we present a simple model of this kind.
We will see that when $\tilde{m}/M=\Lambda_S^2/M^2$ is negligible, 
the model becomes a special version of the MSSM with $\tan\beta=1$, 
and the $h^0$ Higgs boson has vanishing tree-level mass.  
When  $\tilde{m}/M=\Lambda_S^2/M^2$ is sizeable, extra terms 
become important, which in particular make $h^0$ massive.

The superpotential, gauge kinetic functions and 
K\"ahler potential are chosen as
\be W =
\Lambda_S^2 T \ , \;\;\;\;  f_{ab} = {\delta_{ab} \over g_a^2} \left(1
+ 2 {\eta_a \over M} T \right)   \ee 
and
\bea 
K & = &  |T|^2 + |H_1|^2 + |H_2|^2  - {\alpha_t \over 4 M^2}
|T|^4 - {\gamma \over 2 M^4} \left(|H_1|^2 + |H_2|^2 - {v^2\over
2}\right)^2 |T|^2 
\nonumber\\ 
& &  + \left[ |H_1|^2 + |H_2|^2  - ( H_1\cdot H_2 + {\rm h.c.})\right]
\left[ {\beta_{\mu} \over M}(T +\bar{T}) - 
{\alpha \over M^2}|T|^2 \right]\ ,  
\eea
where all parameters are real. 
In this example $W$ is minimal, whereas $K$ is not. In fact, here  
the fields do not correspond to normal coordinates. In principle
we could re-write the model in normal coordinates through 
field redefinitions, but such a coordinate change is not  
convenient in this case. Indeed, the model has been 
specifically designed in the above form in order to allow for a 
simple minimization of $V$, simple VEVs and a simple spectrum. 
In particular, some coefficients in $K$ have been adjusted in such
a way that the minimization of $V$ can be performed exactly, so the 
perturbative procedure sketched in the previous sections 
is not necessary. The basic results can be summarized as follows.

i) The minimum lies at  $\langle T \rangle=0$ and
$\langle H_1^0 \rangle =  \langle H_2^0 \rangle =v/2$,
{\it i.e.} we have $\tan\beta=1$. The metric is canonical at the  
minimum: in particular, the components of $T$ and Higgs 
supermultiplets have no kinetic mixing.

ii) Supersymmetry is broken by the auxiliary component of $T$,  with
$\langle F^T \rangle = -\Lambda_S^2$, whereas  $\langle F^{H_i}
\rangle = 0$ and  $\langle D^a \rangle = 0$. The SUSY-breaking
scale is simply $\sqrt{F}=\Lambda_S$.

iii) The gauge symmetry is broken by the Higgs VEVs, and the $W$ and
$Z$ masses have the usual expressions $M_W^2 = g^2 v^2/4$, $M_Z^2 =
(g^2 + g_Y^2)\ v^2/4$.

iv) In the fermion sector, $\tilde T$ does not mix with the 
other fields and coincides with the goldstino.  The Higgsinos
have mass $\mu = \beta_\mu \tilde{m}$ and the gauginos  have mass $M_a =
\eta_a \tilde{m}$.  The breaking of the electroweak symmetry also
generates mixed gaugino-Higgsino terms as usual.  It is convenient to
use the symmetric and antisymmetric combinations of neutral Higgsinos,
{\it i.e.}~$\tilde{H}_A^0 \equiv  (\tilde{H}_1^0 -
\tilde{H}_2^0)/\sqrt{2}$  and $\tilde{H}_S^0 \equiv (\tilde{H}_1^0 +
\tilde{H}_2^0)/\sqrt{2}$.  The field  $\tilde{H}_S^0$ is a mass
eigenstate and only $\tilde{H}_A^0$  is mixed with gauginos, because of
$\tan\beta=1$.  The neutralino and chargino mass matrices have
standard form, apart from an extra zero eigenvalue corresponding to
$\tilde T$.

v) In the spin-0 sector, $T$-$H$ mixing is absent as well.
The complex field $T$, {\it i.e.} the scalar partner of $\tilde T$, 
has mass $m_T^2= \alpha_t \tilde{m}^2$. The Higgs boson 
spectrum can be summarized as follows: 
\bea
\label{exact0}
G^0 = \rm{Im}(H_1^0+\bar{H}_2^0)  & \rightarrow &  {\rm neutral \;
Goldstone } \\ v + h^0 = \rm{Re}(H_1^0+\bar{H}_2^0)  &
\rightarrow &  m_h^2 =  \gamma v^2 \tilde{m}^2/ M^2 \\ A^0 =
\rm{Im}(H_1^0-\bar{H}_2^0)  & \rightarrow & m_A^2= 2(\alpha +
2\beta_\mu^2) \tilde{m}^2  \\ H^0 = \rm{Re}(H_1^0-\bar{H}_2^0)  &
\rightarrow & m_H^2 = m_A^2 + M_Z^2 \\ G^- = (H_1^- -
\bar{H}_2^-)/\sqrt{2} & \rightarrow & {\rm charged \; Goldstone} \\
H^- =  (H_1^- + \bar{H}_2^-)/\sqrt{2} &  \rightarrow & m^2_{H^{\pm}}=
m_A^2 + M_W^2
\label{exactfin}
\eea where MSSM-type labels have been used.  For $\gamma \tilde{m}^2/M^2
\rightarrow 0$, this would be just the MSSM spectrum for
$\tan\beta=1$. In this limit the electroweak symmetry is
broken along a flat direction and the associated $h^0$ boson 
is massless. The $\gamma$ term has been added 
just to lift this flatness and obtain a nonzero $m_h$,
which can easily be as large as $\sim 500$ GeV.
Notice that the coupling $\gamma \tilde{m}^2/ M^2$ plays 
the role of the coupling $\lambda$ in the SM Higgs potential,
and we can obtain a realistic model if 
$\tilde{m}^2/M^2=\Lambda_S^4/M^4$ is sizeable.

It is straightforward to complete this model (and any other
one) by introducing quark and lepton superfields, so that 
squarks and sleptons obtain ${\cal O}(\tilde{m})$ masses. 
Also notice that we can easily obtain a smooth decoupling limit 
in this model by keeping $v$ fixed and taking $\Lambda_S$ 
and $M$ large with $\Lambda_S/M$ fixed: in this limit 
part of the spectrum
becomes heavy since $\tilde{m}$ becomes large and the
low-energy theory is just the SM, since the light particles are 
only the SM ones, plus the goldstino, which is decoupled.

Although the results we have obtained are exact, it is instructive to
expand the full Higgs potential up to ${\cal O}(H^4)$-terms to make
contact with two-Higgs-doublet models. If we do this, we obtain the 
mass parameters \be
m_1^2=m_2^2=(\alpha+2\beta_\mu^2)\tilde{m}^2
-{1\over 2}\gamma{\tilde{m}^2\over M^2}v^2\  ,\;\;\;\; 
m_3^2=-(\alpha+2\beta_\mu^2)\tilde{m}^2\ , \ee 
where we neglect ${\cal O}(v^4 \tilde{m}^2/M^4)$ terms, 
and quartic couplings 
\bea \lambda_1=\lambda_2&=&{1\over
4}(g^2+g_Y^2)+[\gamma+2(\alpha+2\beta_\mu^2)^2]{\tilde{m}^2\over M^2}\
,\nonumber\\ \lambda_3&=&{1\over
4}(g^2-g_Y^2)+[\gamma+2(\alpha+2\beta_\mu^2)^2]{\tilde{m}^2\over M^2}\
,\nonumber\\ \lambda_4&=&-{1\over
2}g^2+2(\alpha+2\beta_\mu^2)^2{\tilde{m}^2\over M^2}\ ,\nonumber\\
\lambda_5=-\lambda_6=-\lambda_7&=&2(\alpha+2\beta_\mu^2)^2
{\tilde{m}^2\over
M^2}\ ,  \eea 
where we neglect ${\cal O}(v^2 \tilde{m}^2/M^4)$ terms.
This example would thus correspond in Fig.~\ref{thdmbreak}
to point B. If we insert these expressions in the general
formulae for the 2HDM Higgs spectrum (see Appendix A), we
formally recover the results in Eqs.~(\ref{exact0})-(\ref{exactfin}).

We conclude this section by showing an illustrative example 
of goldstino interaction with a Higgs-Higgsino pair.
Although we have not discussed this topic before,
we stress that these couplings are in general present,
{\it i.e.} they are not specific of the model under consideration,
and could be phenomenologically relevant for the 
decay of a Higgsino into a goldstino and a Higgs boson
(see {\it e.g.} \cite{hgh}) or, viceversa,
for the decay of a Higgs boson into a goldstino-Higgsino 
pair (see {\it e.g.} \cite{Djouadi:1997gw}).
We use the above model only to check that such 
couplings have the standard (model-independent) form $\Delta m^2/F$ 
\cite{Fayet:vd}, where $\Delta m^2$ is the mass splitting of the 
fermion-sfermion pair under consideration\footnote{
In the case of mixed states, $\Delta m^2$ is replaced by 
a combination of mass eigenvalues and mixing 
angles.}. To avoid the complications of mixing effects,
we focus on the cubic interactions of the goldstino 
({\it i.e.} $\tilde{T}$) 
with the Higgs boson $h^0$ and the Higgsino $\tilde{H}_S^0$,
which are mass eigenstates and belong to the same
supermultiplet, {\it i.e.} $(H_1^0+H_2^0)/\sqrt{2}$.
The Lagrangian contains both non-derivative and derivative 
interactions of that type. Using the fermion equations of motion
we can write the derivative ones in non-derivative form 
and combine them with the other ones.  
Once this is done, the effective on-shell interaction can be written 
in the simple form
\be - {1 \over \sqrt{2}}{m_h^2 - \mu^2 \over \Lambda_S^2}\,
h^0 \tilde{H}_S^0 \tilde{T} + {\rm h.c.}  \, , 
\ee 
which is the expected result.

\section{Electroweak breaking and two-goldstino interactions}
\setcounter{equation}{0}
\renewcommand{\theequation}{6.\arabic{equation}}

The effective interactions of {\em one} goldstino with a fermion-boson
pair, which are uniquely determined by supercurrent conservation,
can be expressed in terms of the corresponding masses (and
mixing angles) and of the SUSY-breaking scale \cite{Fayet:vd}.
In the last example of the previous section we have checked
that the model-independent form of such couplings is
respected also in the Higgs sector, where electroweak
breaking takes place. 
We devote this section to study the impact of electroweak breaking 
on the effective interactions that involve {\em two} goldstinos. 
We recall that even if the available experimental energy is not 
sufficient to produce the SUSY partners of SM particles, 
SUSY can still be probed in processes involving SM particles 
and two goldstinos. 
Since the corresponding amplitudes are strongly constrained 
by general goldstino properties, by comparison with
experiments one can obtain useful information 
on the SUSY-breaking scale 
\cite{gcoll,Luty:1998np,Brignole:2000wd}.
We would like to study how the coefficients of such 
interactions are affected by electroweak breaking.
To this purpose, we resort to the general framework of sect.~2. 
So our starting point is a general effective theory with 
linearly realized SUSY and $SU(3) \times SU(2) \times U(1)_Y$ 
gauge group\footnote{We also neglect non-singlet terms in $f_{ab}$
and a possible Fayet-Iliopoulos term for $U(1)_Y$,
and assume that $R$-parity is conserved.}.
The chiral supermultiplets include the MSSM ones and 
singlets (in the simplest case, just one $T$ field). 
In the limit of unbroken $SU(2) \times U(1)$,
SUSY can only be broken by the $F$-terms of the singlets:
in this case, the goldstino and its bosonic partners belong
to the singlet sector. Upon $SU(2) \times U(1)$
breaking, SUSY breaking can receive additional
contributions from non-vanishing values of
$\langle F^{H_1^0} \rangle$, $\langle F^{H_2^0} \rangle $, 
$\langle D^{3} \rangle $ and $\langle D^{Y}  \rangle$.
If this is the case, also the neutral Higgsinos and 
gauginos have components along the goldstino.
Moreover, the neutral Higgs bosons and the $Z$ boson
are (partly) bosonic partners of the goldstino.
This implies that such bosons can have non-vanishing
on-shell couplings to goldstino bilinears, as
we will check below. More precisely, in the following 
we will discuss the effective
interactions between two goldstinos and:
{\em i)} a $Z$ boson; {\em ii)} a Higgs boson;
{\em iii)} two SM fermions (leptons or quarks).

We recall that the total amount of SUSY breaking 
is parametrized by $F^2 \equiv \langle V_F + V_D \rangle$,
as in eq.~(\ref{fdef}). 
The indices $i,j,\ldots$ below will run over electrically
neutral chiral supermultiplets, which can have
non-vanishing VEVs in their lowest or auxiliary
components ({\it i.e.} $T$, $H_1^0$ and $H_2^0$,
which will be treated on the same footing).
We also emphasize that, in contrast to our approach
in other sections, throughout our derivations
below we will {\em not} expand the basic functions
in powers of Higgs fields, both for the
sake of generality and for technical convenience\footnote{ 
We will only approximate $\langle D^{3} \rangle $ and 
$\langle D^{Y}  \rangle$ with their lowest order expressions 
after finding general results.}.
 
\subsection{$Z$--goldstino--goldstino}

A connection between the  $Z$-goldstino-goldstino coupling
and non-vanishing electroweak $D$-terms was found in 
\cite{Luty:1998np}, in the framework of non-linearly realized SUSY.
Here we present an alternative derivation
of such a coupling, starting from a general
effective Lagrangian with linearly realized SUSY.
Let us consider the coupling of a generic neutral gauge 
boson\footnote{The symbols $A^a_{\mu}$ and $D^a$ 
correspond here to {\em canonically normalized} fields.} 
$A^a_{\mu}=\{ W^3_{\mu},B_{\mu}\}$ to fermion bilinears 
$\bar{\psi}^{\bar{\imath}} \bar{\sigma}^{\mu} \psi^j$,
where the fermions belong to electrically neutral chiral 
multiplets $(\varphi^i,\psi^i,F^i)$, with $t^a_i$
denoting the weak isospin or the hypercharge. 
After selecting the goldstino components of the fermions
($\psi^i \supset \tilde{G} \langle F^i\rangle/F $), we obtain
\be
\label{agg1}
- {g_a \over F^2} \langle
\bar{F}^{\bar{\imath}} \left( K_{\bar{\imath} j} t^a_j +   
K_{\bar{\imath} j \ell} t^a_{\ell} \varphi^{\ell}\right) F^j \rangle 
A^a_{\mu} \tilde{\bar{G}} \bar{\sigma}^{\mu} \tilde{G}\ ,
\ee
where $g_a \equiv \langle (Re f_a)^{-1} \rangle$ 
is the gauge coupling of $A^a_{\mu}$.
We recall that, upon electroweak breaking, the
goldstino can also have components along neutral gauginos, 
for non-vanishing $\langle D^a \rangle$. However,
such components do not contribute to the coupling
of $A^a_{\mu}$ to goldstino bilinears\footnote{Indeed, the 
interaction $A^a_{\mu} \bar{\lambda}^b \bar{\sigma}^{\mu} \lambda^c$
does not involve neutral gauginos and the interaction
$F_{\mu \nu}^a \lambda^b \sigma^{\mu \nu} \psi^i$ 
cannot contribute because $\tilde{G} \sigma^{\mu \nu} 
\tilde{G}=0$.}. This could give the impression that
the $Z$-goldstino-goldstino coupling is only determined by
$F$-breaking, with electroweak $D$-breaking playing no role.
However, a closer inspection reveals that the
coupling is non zero only if $\langle D^a \rangle \neq 0$.
Indeed, the VEVs $\langle F^i \rangle$ and $\langle D^a \rangle$
are related by the extremum conditions of the scalar
potential. Using these conditions and the constraints from 
gauge invariance we can write the coupling above in
terms of $\langle D^a \rangle$:
\be
\label{agg2}
{1  \over 2 F^2}
\langle D^a \rangle M^2_{ab} \, 
A^b_{\mu} \tilde{\bar{G}} \bar{\sigma}^{\mu} \tilde{G}\ ,
\ee
where $M^2_{ab}$ is the gauge boson mass matrix and 
$\langle D^a \rangle = - g_a \langle K_j t_j^a \varphi^j \rangle$.
Therefore the coupling of a $Z$ boson to a goldstino pair is
\be
\label{zggint}
{\langle D_Z \rangle M^2_Z \over 2 F^2}
Z_{\mu} \tilde{\bar{G}} \bar{\sigma}^{\mu} \tilde{G}\ ,
\ee
where $\langle D_Z \rangle= \langle c_w D^3 -s_w D^Y \rangle 
\simeq - (M_Z^2/g_Z) \cos 2 \beta$ [with
$g_Z \equiv \sqrt{g^2 + g_Y^2} = e/(s_w c_w)$]. The 
associated decay width is
\be
\Gamma(Z \rightarrow \tilde{G}\tilde{G}) 
= {\langle D_Z \rangle^2 M_Z^5 \over 96 \pi F^4 }
\simeq  
\cos^2 2 \beta \left( {200 \, {\rm GeV} \over \sqrt{F}} \right)^8
{\rm MeV}\ ,
\ee
in agreement with \cite{Luty:1998np}. The on-shell
equivalence of the operator (\ref{agg2}) above
to the one found in \cite{Luty:1998np}, {\it i.e.}~${\langle D_a \rangle 
\over 2 F^2 } F^a_{\mu\nu} \partial^{\mu} 
\tilde{\bar{G}} \bar{\sigma}^{\nu} \tilde{G} + {\rm h.c.}$,
can also be checked through the goldstino and gauge boson 
equations of motion. By doing this, in fact, the latter 
operator can easily be converted into the operator (\ref{agg2}).

\subsection{Higgs-goldstino-goldstino}

The coupling of a neutral scalar particle to two (on-shell) 
goldstinos can be derived from the field-dependent
neutral-fermion mass matrix. We can take eq.~(\ref{generalM})
without VEVs, expand the
coefficients of $\lambda^a \lambda^b$,
$\lambda^a \psi^j$, $\psi^i \psi^j$ 
to linear order in the scalar fluctuations
($\delta\varphi^i =  \varphi^i -\langle \varphi^i \rangle$) 
and select the goldstino components of the fermion fields,
eq.~(\ref{gfrag}).
The resulting expression is quite involved:
it depends on $\langle F^i \rangle$, $\langle D^a \rangle$
and several derivatives of $K$, $W$ and $f_{ab}$.
However, by using once again the extremum
conditions of the scalar potential and
gauge invariance, the coefficients of
the scalar-goldstino-goldstino interactions  
can be expressed in terms of the scalar masses 
and $\langle F^i \rangle$. 
The result reads:
\be  
\label{phgg}
{1 \over 2 F^2}\langle F^i \rangle
\left( M^2_{i \bar\jmath} \, \delta\bar{\varphi}^{\bar{\jmath}}
+ M^2_{i j} \, \delta{\varphi}^j \right) 
\tilde{G} \tilde{G} + {\rm h.c.}\ ,
\ee
where $ M^2_{i \bar\jmath} \equiv \langle V_{i \bar\jmath} \rangle$
and $ M^2_{i j} \equiv \langle V_{i j} \rangle$ are
the elements of the scalar mass matrix\footnote{The scalar fields
and the associated masses are not yet canonically normalized 
in (\ref{phgg}). After canonical normalization through 
appropriate use of the K\"ahler metric 
$\langle K_{\bar{\imath} j} \rangle$, the normalized version 
of (\ref{phgg}) can be written in an analogous form.
It could also be written in terms of mass eigenstates
and mixing angles.}. 
Notice the similarity of the boson-goldstino-goldstino 
interactions in (\ref{phgg}) with those in (\ref{agg2}):
in both cases the coefficients are proportional
to the corresponding boson masses, to the VEVs of the
associated auxiliary fields and to $1/F^2$.
In the limit in which SUSY is only broken by the F-term 
$\langle F^T \rangle$ of a singlet superfield $T$ 
and the $T$-scalars have neither kinetic nor mass 
mixing with the Higgses, then only the $T$-scalars couple to 
two on-shell goldstinos and (\ref{phgg}) reduces to known results 
\cite{Casalbuoni:1988sx,Brignole:1996fn}.
Electroweak breaking, however, generically
induces also non-vanishing values of $\langle F^{H_i^0} 
\rangle$ and $T$-$H$ mixing, so also neutral 
Higgs bosons can couple to goldstino bilinears.
The typical size of such couplings is ${\cal O}(v \mu M_H^2/F^2)$,
where $M_H$ denotes the Higgs boson mass. More specific expressions
can be obtained in any given model.
Thus a neutral Higgs boson can decay invisibly into a goldstino pair.
We recall that, in the limiting case of a sizeable invisible width,   
also the branching fractions of the visible channels are
indirectly modified.

\subsection{Goldstino interactions with matter fermions}

We now consider (two goldstino)-(two fermion) effective 
interactions. We recall that, although the leading 
energy- and $F$-dependence of such interactions is completely 
fixed, the presence of (non-universal) model dependent coefficients 
is allowed by general results on non-linearly realized SUSY 
\cite{Brignole:1997pe,Clark:1997aa}. This has also been confirmed
in specific string (brane) constructions \cite{Antoniadis:2001pt}. 
In the framework of effective Lagrangians with linearly 
realized SUSY, a possible source of such parameters
is the presence of $D$-type SUSY breaking 
(besides $F$-type SUSY breaking). Indeed, in this case
the effective (two goldstino)-(two fermion) interactions 
depend on the fermion quantum numbers under the gauge group 
with non vanishing $D$ terms, because of the exchange contribution 
due to the associated massive gauge bosons 
\cite{Fayet:yb,fay,Brignole:2000wd}. 
A physically relevant example of such a situation is precisely 
the case of a SUSY effective Lagrangian with 
gauge group $SU(3)\times SU(2)\times U(1)$ spontaneously broken to 
$SU(3) \times U(1)_{em}$, since the electroweak $D$-terms 
can have non-vanishing VEVs. Therefore, let us consider this case 
in more detail. 

Let $f$ denote a Weyl fermion in the
lepton/quark sector, with isospin $t_f^3$ and electric charge $Q_f$.
The (on-shell) interactions involving two goldstinos and 
two $f$-type fermions arise from three sources: 
sfermion exchange, $Z$ exchange and contact 
interaction\footnote{Since we are interested here in light SM
fermions, we neglect terms of the SUSY effective Lagrangian that
violate the associated chiral symmetries (for instance operators 
that, upon electroweak breaking, generate fermion mass terms
$m_f f f^c$ or left-right sfermion mixing terms).
We recall that, even if such terms are included, low-energy 
cancellations still take place. In this case, the cancellation mechanism 
also involves extra contributions from sfermion exchange 
and contact interactions, as well as additional contributions 
from the exchange of the scalar partners of the goldstino 
\cite{Brignole:1996fn,Brignole:2000wd}. The couplings in 
(\ref{phgg}) are one of the ingredients that lead to such 
cancellations.}.
We recall that the sfermion mass has two contributions
(from $V_F$ and $V_D$):
\be
\tilde{m}^2_f = (\tilde{m}^2_f)_F +  (\tilde{m}^2_f)_D
=\langle \bar{F}^{\bar{\imath}}
(-\log K_{ \bar{f} f})_{\bar{\imath} j} F^j \rangle
- g_Z \langle D_Z \rangle Q^Z_f\ ,
\ee
where $Q^Z_f \equiv t_f^3 - Q_f s_w^2 + 
\langle (\log K_{ \bar{f} f})_j t_j^3 \varphi^j \rangle$
and $K_{ \bar{f} f}$ denotes the K\"ahler metric of
the supermultiplet $(\tilde{f},f)$.
The relevant interaction terms, including the one in 
(\ref{zggint}), are:
\be
\label{gint}
\left[ {\langle D_Z \rangle M^2_Z \over 2 F^2}
\tilde{\bar{G}} \bar{\sigma}^{\mu} \tilde{G}
-g_Z Q^Z_f \bar{f} \bar{\sigma}^{\mu} f \right] Z_{\mu} 
+ {\tilde{m}^2_f \over F}
\left( \tilde{f}^* f \tilde{G} + {\rm h.c.} \right)
-{(\tilde{m}^2_f)_F \over 2 F^2}
(\bar{f} \bar{\sigma}^{\mu} f)
(\tilde{\bar{G}} \bar{\sigma}_{\mu} \tilde{G}) ,
\ee
where all fields are canonically normalized.
An important consequence of the close connection between 
mass spectrum and goldstino couplings, which is manifest 
in eq.~(\ref{gint}), is that the different contributions
to (two goldstino)-(two fermion) interactions cancel 
against each other at zero momentum, as they should.
The first non-vanishing terms in the momentum
expansion ({\it i.e.}~the leading terms for energies smaller 
than $M_Z$ and $\tilde{m}_f$) contain two derivatives
and have the form
\be
\label{ffgg}
- { 1 \over F^2} \left[
(\bar{f} \tilde{\bar{G}} ) \Box (f \tilde{G})
+ {1\over 2} c_f
(\bar{f} \bar{\sigma}^{\mu} f) \Box  
(\tilde{\bar{G}} \bar{\sigma}_{\mu} \tilde{G})   
\right]
\ee
where $c_f$ has the specific value
$c_f= g_Z \langle D_Z \rangle Q^Z_f/ M_Z^2 \simeq 
- (t_f^3 - Q_f s_w^2) \cos 2\beta$. The result in 
(\ref{ffgg}) is consistent with the general form 
allowed by non-linearly realized SUSY\footnote{The 
normalization of $c_f$ in (\ref{ffgg}) is related 
to other parametrizations through the relations 
$c_f -1 = {1 \over 4} \alpha \cite{Brignole:1997pe}= 
- {1 \over 2} C_{ff} \cite{Clark:1997aa} = -C^{(f)} 
\cite{Brignole:2000wd}$.}.
If one is interested in (two goldstino)-(two fermion) 
interactions at higher energies, the local operators 
in (\ref{ffgg}) should be generalized 
to include the full effect of $\tilde{f}$ 
and $Z$ propagators: 
this amounts to replace 
$\Box \rightarrow \Box (1 + \Box/\tilde{m}_f^2)^{-1}$
in the first operator and
$\Box \rightarrow \Box (1 + \Box/M_Z^2)^{-1}$
in the  second operator. 

Let us focus on the process  
$f \bar f \rightarrow  \tilde{G} \tilde{G}$
at $s \ll \tilde{m}^2_f$ and consider the effect
of the $Z$ threshold. The cross-section is
\bea
\label{crffgg}
\sigma(f \bar f \rightarrow \tilde{G}\tilde{G})
& = & 
{ s^3 \over 80 \pi F^4} \left[ 
1 + {5 \over 2}(t_f^3 - Q_f s_w^2) \cos 2\beta \; A(s/M_Z^2)
\right.
\nonumber\\
& & \left.
+ {5 \over 3}(t_f^3 - Q_f s_w^2)^2 \cos^2 2\beta  \; B(s/M_Z^2)
\right]
\eea
where $A(y)\equiv(1-y)B(y)\equiv(1-y)/[(1-y)^2 + \epsilon^2 y^2]$
and $\epsilon=\Gamma_Z/M_Z$ takes into account the finite $Z$ width.
When $\langle D_Z \rangle=0$ ({\it i.e.}~$\cos 2\beta=0$), 
only the first term in eq.~(\ref{crffgg}) [or in (\ref{ffgg})]
is relevant and the cross section reduces to 
$\sigma \simeq s^3/(80 \pi F^4)$ \cite{Brignole:1997pe}. 
For $\langle D_Z \rangle \neq 0$, however, this simple result only holds 
above the $Z$ threshold, {\it i.e.}~for $M_Z^2 \ll s \ll \tilde{m}^2_f$, 
where the second and third terms in eq.~(\ref{crffgg}) are suppressed 
[$A(y)\sim -1/y$ and $B(y)\sim 1/y^2$ for $y\gg1$]. On the other hand,
such terms become dominant in the resonance region.
Below resonance, all three terms in eq.~(\ref{crffgg})
contribute with comparable weight [$A(0)=B(0)=1$].   
It is straightfoward to extend these results to 
a $SM$ fermion ${\cal F}$ with both helicity components,
{\it i.e.}~$SU(2)$ doublet component $f$ and singlet component 
$\bar{f^c}$. 
The unpolarized cross section for ${\cal F} \bar{\cal F} 
\rightarrow \tilde{G}\tilde{G}$ is easily inferred from 
eq.~(\ref{crffgg}):
\bea
\label{uffgg}
\sigma_{\rm unp}({\cal F} \bar{\cal F} \rightarrow \tilde{G}\tilde{G})
& = & 
{ s^3 \over N \, 160 \pi F^4} \left[ 
1 + {5 \over 4}\, t_f^3 \cos 2\beta \; A(s/M_Z^2)
\right.
\nonumber\\
& & \left.
+ {5 \over 6}\left(
{1 \over 4} -2\, t_f^3 Q_f s_w^2 + 2\,  Q_f^2 s_w^4 \right)
\cos^2 2\beta \; B(s/M_Z^2) 
\right]
\eea
where $N=1 \, (3)$ for charged leptons (quarks).
We also recall that, in order to obtain a visible signal
at colliders, the goldstino pair should be accompanied
for instance by a photon or a gluon, as in $e^+ e^- \rightarrow 
\tilde{G}\tilde{G} \gamma$, $Q \bar{Q} \rightarrow \tilde{G}\tilde{G}
\gamma$, $Q \bar{Q} \rightarrow \tilde{G}\tilde{G} g$
(see {\it e.g.}~\cite{gcoll}).
As an alternative to a full computation, 
approximate expressions for the cross section of such 
five-particle processes can be obtained by convoluting the 
above four-particle cross section with the radiator functions 
that describe initial state radiation. Then the kinematical variable 
$s$ in eq.~(\ref{uffgg}) would be related to the analogous quantity $S$ 
of the five-particle process through $s=S(1-x)$, where $x$ is the 
energy fraction taken away by the photon or the gluon.

\section{Summary and conclusions}

In recent years there has been an intense activity on supersymmetric
models in which the scales of SUSY breaking ($\sqrt{F}$) and
mediation ($M$) are close to the electroweak scale.
These include models of extra dimensions (warped or not)
with low fundamental scale and, more generally,
scenarios in which the low-energy supersymmetric effective
theory is obtained by integrating out physics at energy scales
not far from the TeV scale. 
In all these cases the usual MSSM, where
the effects of SUSY breaking in the observable sector are encoded in a
set of soft SUSY-breaking terms of size ${\cal O}(F/M)$, may not 
give an accurate enough effective description. Additional effects 
can be relevant, in particular 
interactions of the goldstino sector with the observable sector 
and non-negligible contributions to `hard-breaking' terms,
such as ${\cal O}(F^2/M^4)$ contributions to quartic Higgs couplings. 
In fact, the latter contributions can compete with (and may take over) 
the usual ($D$-term induced) MSSM quartic Higgs couplings, giving rise 
to a quite unconventional Higgs sector phenomenology, as already
observed in \cite{Polonsky}. 
The main purpose of this paper has been to study in detail
the latter aspect, {\it i.e.}~to perform a general analysis 
of the Higgs sector and the breaking of SUSY and electroweak 
symmetry in this type of models. 
To do this, we have used a model-independent approach based 
on a general effective Lagrangian, in which the MSSM superfields
are effectively coupled to a singlet superfield, assumed to be
the main source of SUSY breaking. 
Our main results can be summarized as follows:

\begin{itemize}

\item Rather than the usual MSSM potential, the Higgs potential
resembles that of a two-Higgs-doublet model (2HDM), where the
quadratic and quartic couplings can be traced back to the original
couplings in the effective superpotential and K\"ahler potential. 
However, there are still some differences, {\it e.g.}~the presence of
derivative couplings besides the non-derivative ones described
by the scalar potential. Moreover, the scalar sector also contains 
an extra complex degree of freedom, which comes from the
singlet supermultiplet. This scalar field can have 
non-negligible interactions with the Higgs fields, and 
could also mix with them as a result of electroweak breaking.

\item The presence of extra quartic couplings that may be larger
than the usual ones opens novel opportunities for electroweak
breaking. The breaking process is effectively triggered 
at tree-level and presents important differences with the usual 
radiative mechanism. Electroweak breaking can occur in a much
wider region of parameter space, {\it i.e.} for values of the 
low-energy mass parameters that are normally forbidden.
For instance, $m_{H_1}^2$ and $m_{H_2}^2$ are allowed to be
both negative. Another unconventional situation, now allowed, is 
the case in which $m_{H_1}^2$ and $m_{H_2}^2$ are equal and
positive, and electroweak breaking is driven by 
$m_3^2 \ H_1\cdot H_2$. This breaking is natural, since 
the latter term is the only off-diagonal bilinear coupling 
among MSSM fields (with R-parity conserved), so
$SU(2)_L\times U(1)_Y$ is the only symmetry
that can be broken when all scalar masses are positive.
A further advantage of the extra quartic couplings is
that their presence may reduce the amount of tuning 
necessary to get the correct Higgs VEVs.

\item The spectrum of the Higgs sector is also dramatically changed,
and the usual MSSM mass relations are easily violated.
In particular, the new quartic couplings allow the lightest 
Higgs field to be much heavier ($\lsim 500$ GeV) than in usual 
supersymmetric scenarios. Moreover, this field could have a 
substantial singlet component, modifying its properties.

\item Departures from the usual MSSM results also appear
in the chargino and neutralino mass matrices, where 
the effective operators induce some corrections after 
electroweak breaking. Moreover, the neutralino sector
also includes the goldstino. This is a singlet in the limit 
of unbroken electroweak symmetry, but generically 
(although not necessarily) also acquires Higgsino and 
gaugino components after electroweak breaking.

\item After giving a general derivation and discussion of the 
above properties, we have illustrated them in 
two simple examples, analyzing in each case the Higgs potential, 
the electroweak breaking process, the Higgs masses
and the neutralino and chargino spectra.

\item 
Finally, we have analysed the role of electroweak 
breaking in processes in which SM particles could emit a 
goldstino pair, such as fermion-antifermion annihilations
and the invisible decays of $Z$ and Higgs bosons.
We recall that such processes may offer an important window 
to SUSY, especially if other superparticles are not experimentally 
accessible. 

\end{itemize}

In conclusion, it is clear that many features of the 
conventional MSSM Higgs sector and related ones can be significantly 
changed in scenarios with low-scale SUSY breaking (examples of this 
are the mechanism of the electroweak breaking and the mass of the 
lightest Higgs). This potentially offers new ways to overcome traditional
difficulties  of the MSSM as well as new prospects for the detection
of SUSY in future experiments.

\section*{Acknowledgments}
This work was partially supported by the European Programmes
HPRN-CT-2000-00149 (Collider Physics) and HPRN-CT-2000-00148 (Across
the Energy Frontier).
J.R.E. thanks the CERN TH-Division for financial support during the       
final stages of this work.

\section*{Appendix A}
\setcounter{equation}{0}
\renewcommand{\theequation}{A.\arabic{equation}}

This Appendix deals with a subclass of quartic two-Higgs-doublet 
potentials: those which have real parameters and are
invariant under a symmetry that exchanges the two doublets.
The mass parameters and the quartic couplings of such a
potential are subject to the restrictions $m_1^2=m_2^2$,
$\lambda_1=\lambda_2$ and $\lambda_6=\lambda_7$, 
{\it i.e.}~the potential has the form
\bea
V(H_1,H_2) & = & m_1^2 (|H_1|^2 + |H_2|^2) + 
m_3^2 (H_1\cdot H_2 + {\rm h.c.})
\nonumber\\
& + &
{1\over 2} \lambda_1 (|H_1|^4 + |H_2|^4)  + 
\lambda_3 |H_1|^2 |H_2|^2 + \lambda_4 |H_1\cdot H_2|^2
\nonumber\\
& + &
{1 \over 2} \lambda_5 \left[ (H_1\cdot H_2)^2  + {\rm h.c.} \right] 
+ \lambda_6 (|H_1|^2 + |H_2|^2)( H_1\cdot H_2 + {\rm h.c.}) .
\label{symV}
\eea
A special case of symmetric potentials are those 
invariant under $SU(2)_L\times SU(2)_R$. Such
potentials only depend on the quantities 
$|H_1|^2+|H_2|^2$ and $H_1\cdot H_2$, so the
further condition $\lambda_1=\lambda_3$ holds 
for them.

If a symmetric potential of the form (\ref{symV}) admits a 
minimum with nonvanishing 
Higgs VEVs, such a minimum could either preserve 
($|\tan\beta|=1$) or break spontaneously ($|\tan\beta| \neq 1$) 
the discrete symmetry that exchanges the two doublets.
We will now present the conditions under which the former 
or the latter case is realized, and give explicit formulae
for the Higgs boson masses\footnote{
Such formulae do not include the corrections from 
kinetic normalization, which should be added 
if required. Notice, however, that these corrections
are higher order effects for Higgs masses that
are ${\cal O}(v^2)$ at leading order.}. 
We anticipate that, in the special case of $SU(2)_L\times SU(2)_R$
invariant potentials, only the case $|\tan\beta |=1$ can be realized.
For later convenience, we introduce the 
following abbreviation: $\tilde\lambda \equiv (1/2)
(-\lambda_1+\lambda_3+\lambda_4+\lambda_5)$.

\vskip 1 cm
\begin{center}
{\bf \large Minima with $|\tan\beta|=1$}
\end{center}
The conditions to have a minimum with $\tan\beta=s=\pm1$ 
are:
\bea
m_1^2 + s m_3^2 & < & 0\ ,
\\
\lambda_1 + \tilde\lambda + 2 s \lambda_6 & > & 0\ ,
\\
(\tilde\lambda + s\lambda_6) m_1^2 & > & 
(\lambda_1+ s\lambda_6) s m_3^2 \ ,
\\ 
(\lambda_5 + s\lambda_6) m_1^2  & > & 
(\lambda_1 + \tilde\lambda - \lambda_5 + s\lambda_6) s m_3^2\ ,
\\
(\lambda_4 + \lambda_5 + 2 s\lambda_6) m_1^2  & > & 
(\lambda_1 + \lambda_3 + 2 s\lambda_6) s m_3^2 \ .
\eea
The value of $v^2 \equiv 2 \langle |H_1^0|^2 +  
|H_2^0|^2 \rangle$ is
\be
v^2 = { -2(m_1^2 + s m_3^2) \over 
\lambda_1 + \tilde\lambda + 2 s \lambda_6 }\ .
\ee
The mass of the Higgs field along the VEV direction is
\be
m_h^2 = (\lambda_1 + \tilde\lambda + 2 s \lambda_6) \, v^2\ .
\ee
The remaining Higgs boson masses are
\bea
m_H^2 & = & 
2 m_1^2 + (\lambda_1+ s\lambda_6) v^2 \, = \, 
- 2 s m_3^2 - (\tilde\lambda + s\lambda_6) \, v^2\ ,
\\
m_A^2 & = &  
2 m_1^2 + (\lambda_1+\tilde\lambda -\lambda_5+ s\lambda_6)\, v^2 
\, = \, 
- 2 s m_3^2 - (\lambda_5 + s\lambda_6)\, v^2\ ,
\\
m_{H^{\pm}}^2 & = & 
2 m_1^2 + \left[ {1 \over 2}(\lambda_1+\lambda_3)
+ s\lambda_6 \right] v^2 
 =  
- 2 s m_3^2 -\left[ {1 \over 2}(\lambda_4 + \lambda_5) 
+ s\lambda_6 \right] v^2.
\eea

\vskip 1 cm
\begin{center}
{\bf \large Minima with $|\tan\beta| \neq 1$}
\end{center}
The conditions to have a minimum with $|\tan\beta| \neq 1$ 
are:
\bea
m_1^2 - |m_3^2| & < &  0\ ,
\\
\lambda_3 > \lambda_1 & > &  0\ ,
\\
\lambda_1 \tilde\lambda & > & \lambda_6^2\ ,
\\
\tilde\lambda & > & \lambda_5\ ,
\\
\lambda_6 m_3^2 - \tilde\lambda m_1^2 & > & 
|\lambda_6 m_1^2 - \lambda_1 m_3^2|\ .
\eea
The values of $v^2 \equiv 2 \langle |H_1^0|^2 +  
|H_2^0|^2 \rangle$ and $\tan\beta \equiv 
\langle H_2^0/H_1^0 \rangle$ are determined by
\be
v^2 = 
{2(\lambda_6 m_3^2 - \tilde\lambda m_1^2)\over
\tilde\lambda \lambda_1 - \lambda_6^2}\ ,
\;\;\;
\sin2\beta={\lambda_6 m_1^2-\lambda_1m_3^2\over 
\lambda_6 m_3^2-\tilde\lambda m_1^2}\ .
\ee
The CP-even Higgs mass matrix, projected on the VEV direction 
($h_\parallel$) and on the orthogonal one ($h_\perp$), reads:
\bea
\langle h_\parallel | {\cal M}^2 | h_\parallel \rangle
& = &
(\lambda_1 + \tilde\lambda \sin^2 2\beta
+ 2 \lambda_6 \sin2\beta) \, v^2\ ,
\\
\langle h_\perp | {\cal M}^2 | h_\perp \rangle
& = &
(\tilde\lambda \cos^2 2\beta) \, v^2\ ,
\\
\langle h_\parallel | {\cal M}^2 | h_\perp \rangle
& = &
(\tilde\lambda \sin 2\beta + \lambda_6) \cos 2\beta \, v^2 \ .
\eea
These imply the following bounds on the masses of the CP-even Higgs 
bosons:
\bea
&&{\mathrm min}\{m_h^2,m_H^2\}\leq\langle h_\parallel | {\cal M}^2 | 
h_\parallel 
\rangle\leq {\mathrm max}\{m_h^2,m_H^2\}\ ,\\
&&{\mathrm min}\{m_h^2,m_H^2\}\leq\langle h_\perp | {\cal M}^2 | h_\perp 
\rangle\leq {\mathrm max}\{m_h^2,m_H^2\}\ .
\eea
The CP-odd and charged Higgs masses are
\bea
m_A^2 & = & 
{1 \over 2}(-\lambda_1+\lambda_3+\lambda_4-\lambda_5) v^2
\, = \,
(\tilde\lambda - \lambda_5)\, v^2\ ,
\\
m_{H^{\pm}}^2 & = & {1 \over 2} (-\lambda_1 + \lambda_3) \, v^2
\, = \,
\left[ \tilde\lambda - {1 \over 2}(\lambda_4 + \lambda_5) 
\right] v^2\ .
\eea

In particular notice that, in order to have a minimum with 
$|\tan\beta| \neq 1$ and positive $m_{H^{\pm}}^2$,
a symmetric potential has to fulfil the condition 
$\lambda_3-\lambda_1 >0 $.
This is not satisfied by $SU(2)_L\times SU(2)_R$
invariant potentials: in this case a non-trivial minimum 
necessarily has $|\tan\beta|=1$.
The same conclusion holds in those supersymmetric models
in which $SU(2)_L\times SU(2)_R$ is preserved by 
the K\"ahler potential (before inserting the $U(1)_Y$
vector superfield) and is only broken by the hypercharge 
coupling $g'$: in such a case $\lambda_3-\lambda_1 = 
- g'^2/2 < 0 $, so a non-trivial minimum necessarily 
has $|\tan\beta|=1$.

%%%%%%%%%%%%%%%%%%%%%%%%%%%%%%%%%%%%%%%%%%%%%%%%%%%%%%%%%%%%%%%%%%%

%
\end{document}